# Unveiling ADP-binding sites and channels in respiratory complexes: Validation of Murburn concept as a holistic explanation for oxidative phosphorylation

*Kelath Murali Manoj\* & Abhinav Parashar*

\*Corresponding author- Satyamjayatu: The Science and Ethics Foundation, Kulappully, Shoranur-2, Palakkad District, Kerala State – 679122, India. Email: satyamjayatu@yahoo.com

**Summary**

Mitochondrial oxidative phosphorylation (mOxPhos) makes ATP, the energy currency of life. Chemiosmosis, a "proton-centric" mechanism, advocates that Complex V harnesses a trans-membrane potential (TMP) for ATP synthesis. This perception of cellular respiration requires oxygen to stay tethered at Complex IV (an association inhibited by cyanide) and diffusible/reactive oxygen species (DROS) are considered wasteful/toxic products. With new mechanistic insights on heme/flavin enzymes, an "oxygen/DROS-centric" explanation (called murburn concept) was recently proposed for mOxPhos. In this mechanism- TMP is not directly harnessed, protons are a rate-limiting reactant and DROS within matrix serve as the chemical coupling agent, directly linking NADH oxidation with ATP synthesis. We report multiple ADP-binding sites and solvent-accessible DROS-channels in respiratory proteins, which validate the oxygen/DROS-centric power generation (ATP synthesis) system in mOxPhos. Since cyanide's heme-binding $K_d$ is high (~mM), low doses (~μM) of cyanide is lethal because it disrupts DROS dynamics in mOxPhos.

## Graphical Abstract:

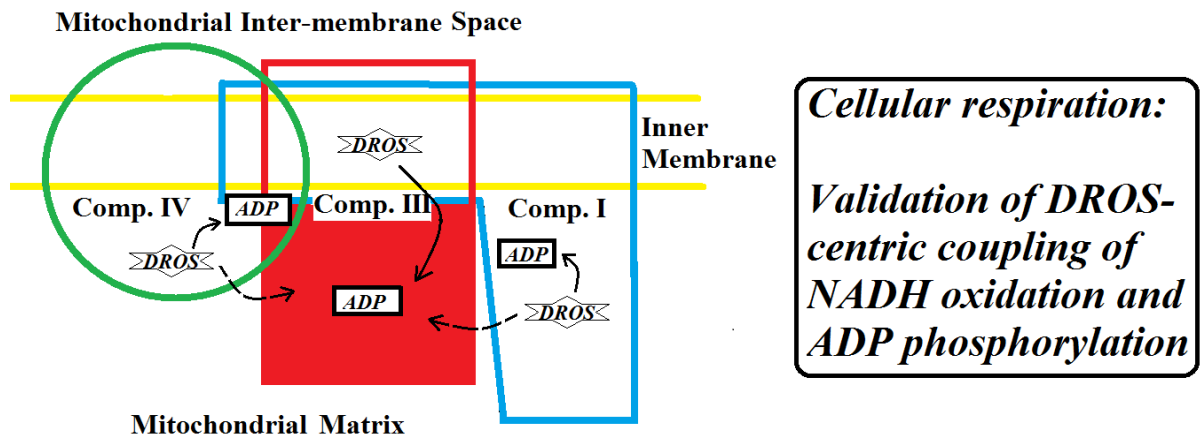

## Highlights:

* Functions of membrane-embedded respiratory (super)complexes are revisited

* Till date, Complexes I-IV are allotted roles in electron transport, not ATP synthesis

* ADP-binding sites and channels/cavities' roles in Complexes I-IV are unveiled

* Kinetics and energetics arguments discount "proton-centric" rotary ATP synthesis

* Findings/structures agree with "oxygen-centric, decentralized" ADP-Pi activation

## Introduction

ATP, the chemical energy-currency of life, is synthesized *in situ* primarily via a metabolic routine called mitochondrial oxidative phosphorylation (mOxPhos). The Mitchell-Boyer explanation (Boyer, 1997; Mitchell, 1961), as acclaimed/detailed in popular biochemistry textbooks (Berg et al., 2002; Lehninger et al., 2004; Voet and Voet, 2011), entails an elaborate "proton-centric" metabolic scheme involving "Rotary ATP synthesis (RAS) – Chemiosmosis principle – Proton pumps – Electron Transport Chain (ETC)" (henceforth collectively anagrammed as RCPE). In this view, diffusible/reactive oxygen species (DROS) are considered toxic/undesired side-products. Further, there is no direct "chemical reaction" connectivity between NADH/succinate oxidation (by Complexes I through IV, on the mitochondrial membrane) and ATP synthesis (by Complex V, in the matrix). In the past, several seasoned researchers have independently questioned many aspects of the RCPE paradigm, even after it was accorded global recognition (Berden, 2003; Ling, 1981; Nałęcz, 1986; Nath, 2010a, b; Slater, 1987; Wainio, 1985). Recently, the crystal structures of mammalian mitochondrial membrane molecular assemblies have provided a wealth of new information (Fiedorczuk et al., 2016; Gu et al., 2016; Stroud et al., 2016; Wu et al., 2016; Zhu et al., 2016). The insights from these works challenged long-held assumptions, such as- (i) structural/functional connectivity of purported proton-pumping domains with electron transport routes & (ii) the functional relevance of Q-cycle. As these components/concepts were integral to the RCPE mechanism (Mitchell, 1975; Zhang et al., 1998), new explanations were mandated in the area.

On a parallel note, two decades' investigations into the role(s) of diffusible species and reactions mediated outside the "active-site" in heme/flavin enzyme systems (Andrew et al., 2011; Gade et al., 2012; Gideon et al., 2012; Manoj, 2006; Manoj et al., 2010a; Manoj et al., 2010b; Manoj et al., 2016a; Manoj and Hager, 2001; Manoj and Hager, 2008; Manoj et al., 2016b; Manoj et al., 2016c; Parashar et al., 2014a; Parashar et al., 2018; Parashar and Manoj, 2012; Parashar et al., 2014b; Venkatachalam et al., 2016) afforded a novel oxygen/DROS-centric mechanistic scheme for explaining select redox processes. This new mechanism was recently called murburn concept (derived from "*mur*ed *burn*ing", connoting a mild and unrestricted but confined and stochastic downhill scheme involving oxygen-centred radicals) (Manoj, 2017; Manoj et al., 2016a; Manoj et al., 2016b; Manoj et al., 2016c; Parashar et al., 2018; Venkatachalam et al., 2016). Murburn concept reasoned long-standing conundrums of microsomal xenobiotic metabolism (mXM) (Manoj et al., 2016c) and maverick physiological dose responses (Parashar et al., 2018). Since the structure-function logic of mXM and mOxPhos systems were quite similar, Manoj had recently proposed a murburn mechanism for mOxPhos (Manoj, 2017). The new proposal accounted for the paucity of protons within the mitochondria, advocated obligatory/constructive roles for DROS and predicted alternate functions for respiratory complexes. In this context, the structure-function features of respiratory complexes are re-investigated herein.

## Materials & Methods

Unlike the proton-centric chemiosmosis hypothesis, murburn concept advocates an oxygen/DROS-centric paradigm. If the oxygen-centric scheme was relevant in mOxPhos, we would expect ADP-binding sites on the complexes, oxygen/DROS to have access to the redox centres within complexes and tunnels/channels in the vicinity of DROS production sites. So, the respiratory protein complexes were screened/analyzed for these features (within the matrix-ward protrusions accessible outside the trans-membrane region) using PyMOL 1.3 and 2.1.1 (DeLano, 2002), AutoDock 4.2 (Morris et al., 2009), Chimera 1.12 (Pettersen et al., 2004), POCASA (online tool) (Yu et al., 2010) and CAVER 3.0 (Chovancova et al., 2012). The relevant legends to figures and Item 1, Supplementary Information file provide details of the proteins, controls and protocols used. Other justifications are available from our earlier publications (Manoj et al., 2016b; Manoj et al., 2016c; Parashar et al., 2018; Parashar et al., 2014b; Venkatachalam et al., 2016).

**Results**

***Exploring putative ADP-binding sites on the respiratory (super)complexes:***
Complexes I through V exhibited ADP-binding sites, which are depicted in Figures 1 & 2. It was noted that the *in silico* $K_d$ values (calculated from the conformer affording lowest binding energy term) of protein Complexes I through V and the overlapping regions within respirasome [I(III)$_2$IV] ranged from $10^{-5}$ M to $10^{-7}$ M, with some Complexes showing multiple binding clusters. *In toto*, the respirasome structure had 18 ADP binding sites, with 8, 4 and 2 sites each within the matrix-ward projections of Complexes I, III and IV respectively. Complex II exhibited a single ADP binding site. The known ADP binder, Complex V, gave two ADP-binding clusters within an α-β dimer, averaging (*in silico*) $K_d$ of ~$10^{-5}$ M, which agrees with textbook values. For more visuals/details, please refer Item 1, Supplementary Information.

*Figure 1: ADP-binding sites depicted on respirasome structure. The sites were detected individually/indpendently in each complex and the overlapping regions of the respirasome structure. The binding sites of ADP below the 3 nm transmembrane region are encircled (except one – inside a hexagon to demarcate the posterior cluster site). Sites depicted here are representative of only one monomer, in case of Complex III and IV.*

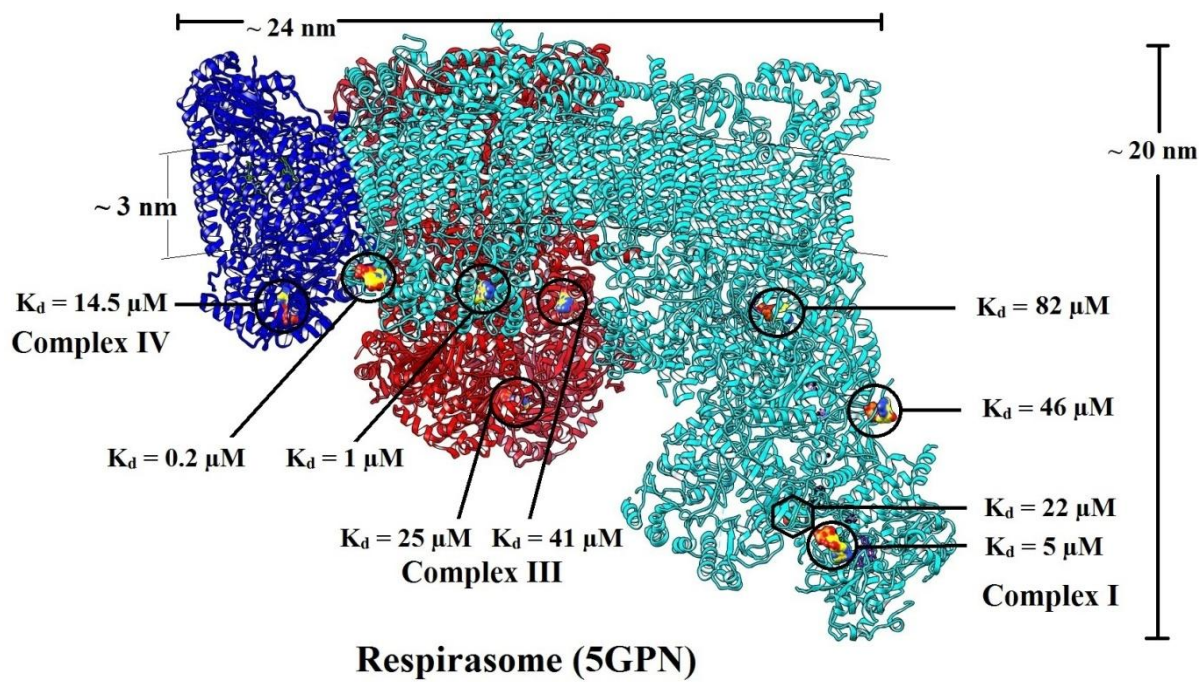

*Figure 2: ADP binding sites on Complexes II (3SFD) and V (1H8E). ADP binding sites (encircled) as detected upon in silico docking demonstrated only one cluster in Complex II, while it showed two sites in Complex V. Only F1 subunit of Complex V is depicted here with alternate orange and pink shades for α and β subunits and the γ stalk in wheat.*

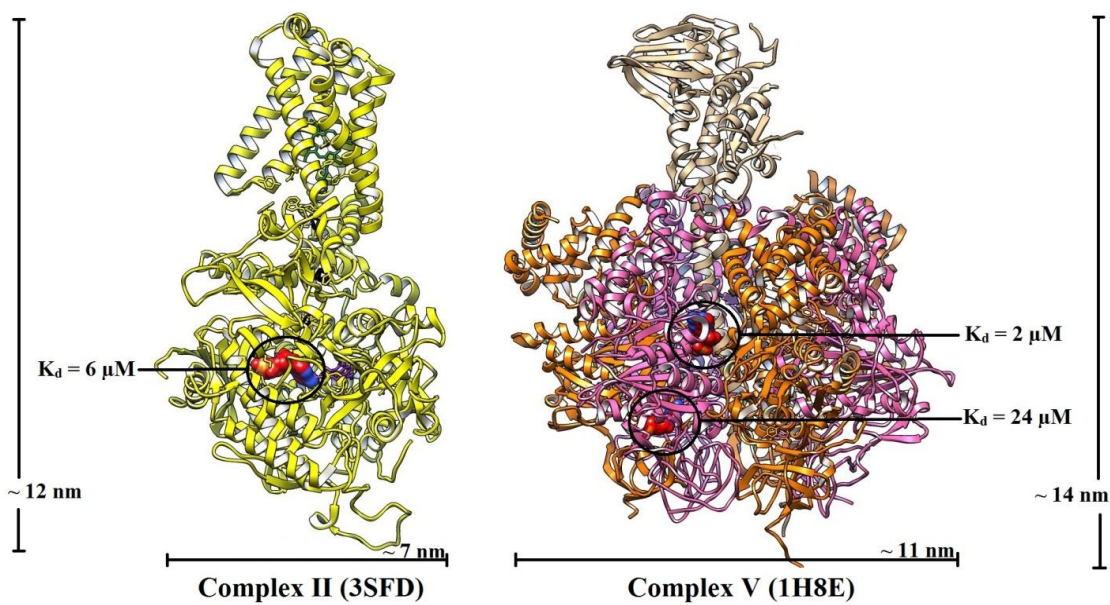

*Tracing $O_2$/DROS channels in the respiratory (super)complexes:*

Figures 3, 4, 5 and 6 detail the cavity locations, solvent-access to redox centres and the bindings sites of ADP on Complexes I, II, III & IV respectively. Complex I shows an extended cavity along the "wired series" of FeS centers starting from the flavin-containing distal end, all the way up to the "CoQ-binding site" near the trans-membrane region. This cavity is connected to the solvent through several channels, all opening towards the direction of its trans-membrane anchor (and towards Complex III's matrix-ward projection within the respirasome structure). At least three of the ADP-binding sites pose favorable loci for an attack by the DROS that could emerge from the channels. Complex II has one solvent-accessible ADP binding site, which is also connected to the flavin and the first Fe-S center via the cavity within the bulbous matrix-ward projection. Complex III, whose matrix-ward projection's function was hitherto unknown, shows a highly "open" dimeric structure, with huge cavities within the trans-membrane and matrix-ward domains. The heme of Cyt $b_H$ would be accessible to oxygen/DROS/CoQ from the trans-membrane and matrix sides. Therefore, it is clear that this redox centre cannot be involved in the "intricate" Q-cycle. The cavity from this heme-centre is connected to the ADP binding sites on the matrix-ward projection of the apoprotein. The ADP-binding site is located adjacent to the matrix-ward channel of heme $a_3$ ($Cu_B$) within Complex IV. Kindly refer Item 1, Supplementary Information file for more visuals/details.

*Figure 3: Cavity/solvent accessible channels and ADP binding sites within Complex I (5LDW). The solvent channels present in the matrix-ward protrusion of Complex I, generated through POCASA 1.1 with a probe radius of 2Å, is demonstrated in blue. In the highlighted zone, the continuity of the solvent channel is clearly seen with some of the Fe-S clusters. Also, the flavin is fully accessible to the cavity.*

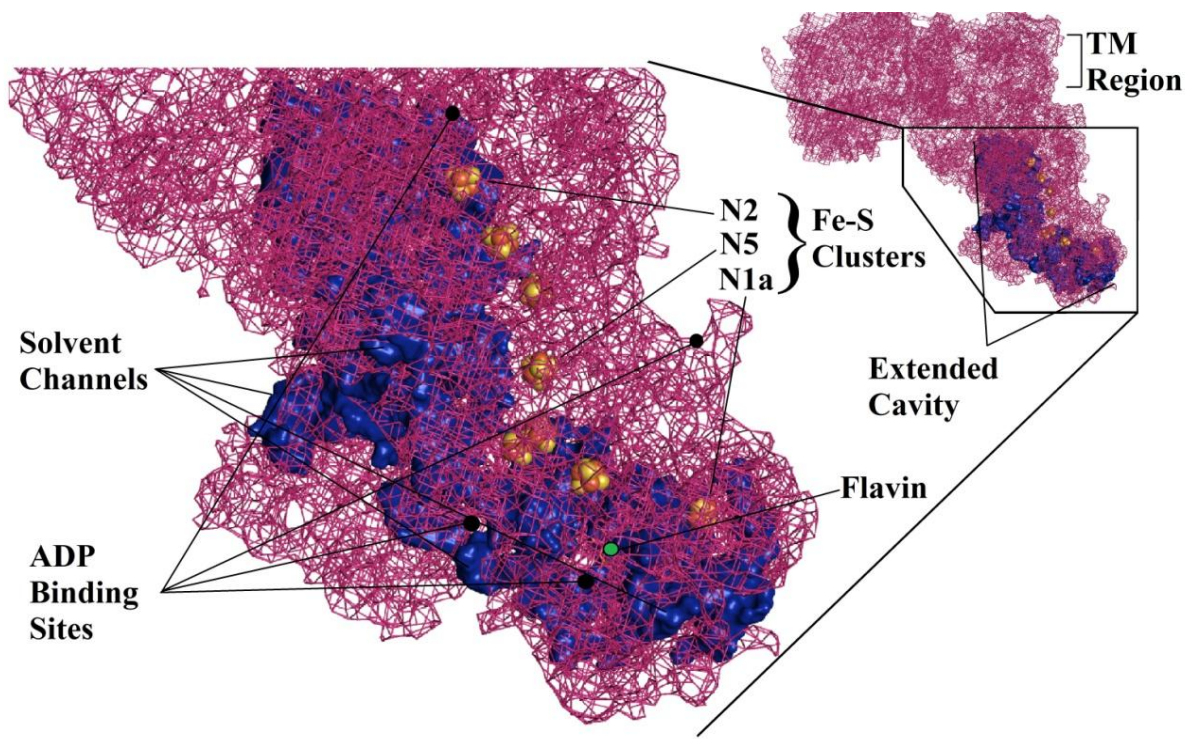

*Figure 4: Cavity/solvent accessible channels and ADP binding sites within Complex II (3SFD).* ADP binding site on Complex II is seen at the entry site of the channel/cavity (generated through POCASA 1.1 with a probe radius of 2Å) is presented with a "mesh/grid view". The first Fe-S center and the flavin are accessible within the same cavity.

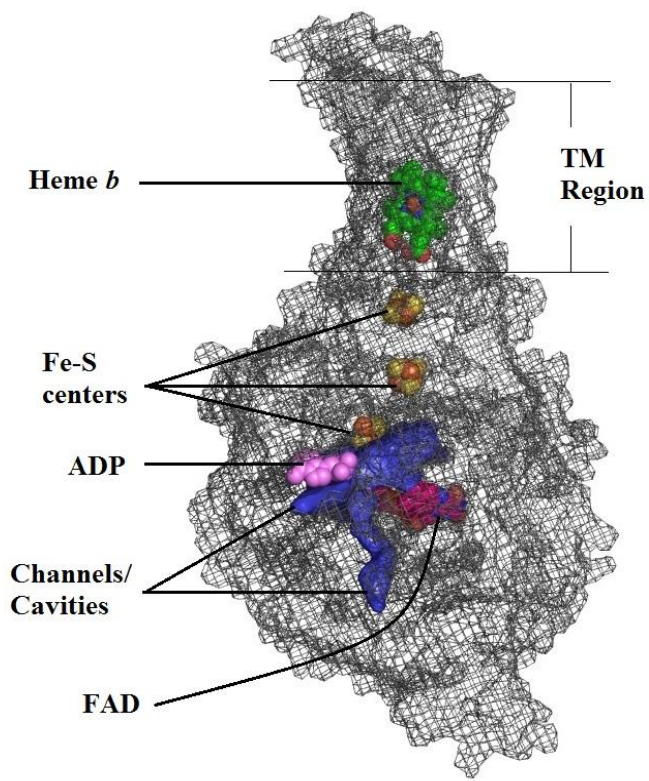

*Figure 5: ADP binding sites on the extensively "open" structure of Complex III (2A06).*
*The transparent surface view of whole Complex III with demarcated monomers (one colored gray and other blue) demonstrates the position of heme-centers, Fe-S centers, ADP binding cluster sites (shown for only one monomer) and channels/cavities. It is also evident from the image that heme $b_H$ is at the verge of transmembrane and has a channel opening into the 'big cavity' between the dimers (Figure A7, Item 1, Supplementary Information).*

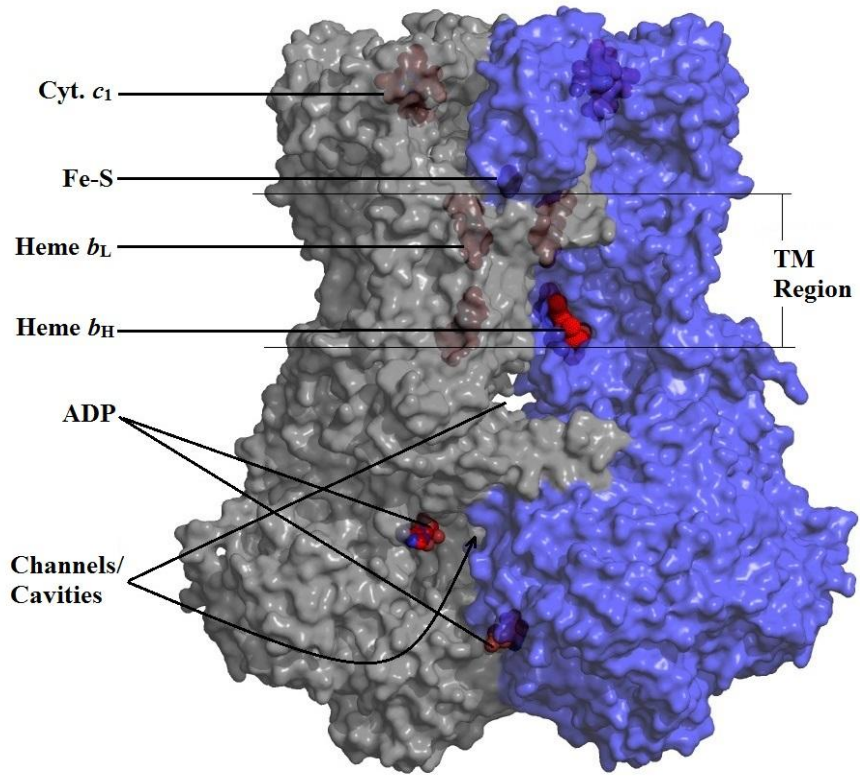

*Figure 6: Location of channels and ADP binding sites on Complex IV (1OCC).* *The long channel shown in blue (result generated using POCASA 1.1 with probe radius of 2Å), indicates good connectivity between $Cu_B$ – Heme $a_3$ and ADP binding site. Only one binding cluster site is present in one monomer, as depicted here.*

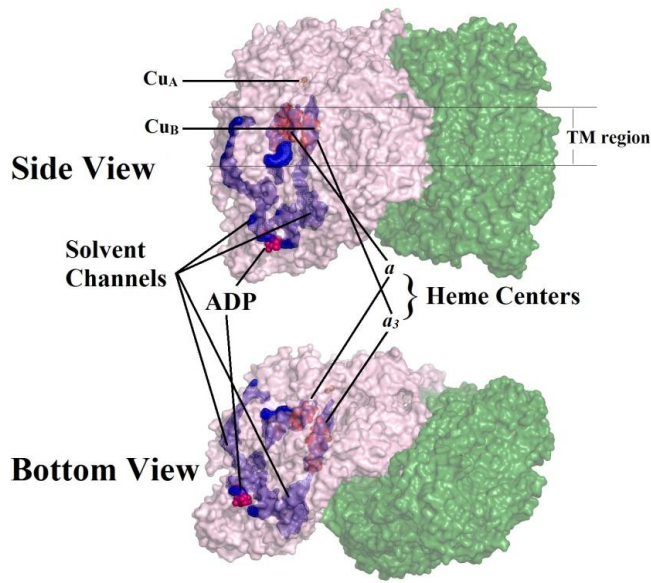

*Quantitative mechanistic analysis of physiological functioning of Complex V:*

In respiring mitochondria, concentrations of Pi, ADP and $H^+$ range at $10^{-2}$ M, $10^{-4}$ M, and $10^{-7}$ M whereas ATP and water are at $\leq 10^{-3}$ M and $> 10^1$ M respectively. Further, Complex V's affinities for "phosphate species" are in the order of Pi ($K_d \sim 10^{-2}$ M) < ADP ($K_d \sim 10^{-5}$ M) << ATP ($K_d \leq 10^{-12}$ M). Let's assume that the physiological reaction is: ADP + Pi + $H^+$ → ATP + $H_2O$. Discounting "highly deterministic enzyme mechanics", the effective activities (with water's and protons' activities taken as unity, a treatment favouring esterification) would imply that Complex V would catalyze the hydrolysis reaction over esterification by a factor of $10^9$! Seen in another way- assuming a diffusion-limited second-order on-rate of $10^8$ $M^{-1}s^{-1}$ for all the interacting entities, the off-rates on $F_1$ would be approximately $10^6$ $s^{-1}$, $10^3$ $s^{-1}$ and $\leq 10^{-4}$ $s^{-1}$ for Pi, ADP and ATP respectively. Then, the physiological ATP synthesis (observed to be $\sim 10^2$-$10^3$ $s^{-1}$) cannot originate from Complex V because the two substrates "bounce off" the enzyme more often (do not "stick on" to get converted) and because the substrate ADP is available less than ATP. Regardless of the way Complex V achieves the transition state (ADP.Pi-E-ATP; from the left via ADP.Pi or from the right via ATP), if the enzyme has high affinity for ATP, it would let go off predominantly ADP and Pi. So, the enzyme would hydrolyze, predominantly. If we consider Complex V as an enzyme which cycles with high affinity for both ATP and ADP (the latter achieved by the γ shaft pushing ATP out, enabling ADP binding) within a single pot, it still cannot give directional synthesis of ATP alone (particularly when ATP is preponderant in milieu!). The fact that purified Complex V is primarily a demonstrable ATPase consolidates the arguments presented herein. Further, it makes little evolutionary or metabolic purpose that nature would employ the same multi-phasic enzyme for both hydrolytic and esterification functions.

The bisubstrate enzyme creatine kinase (catalyzing the reaction: Phosphocreatine + ADP ↔ Creatine + ATP) can be taken as an equilibrium-driven/achieving enzyme similar to Complex V. Table 1 presents an excerpt of experimental kinetics/equilibrium constants, taken from a review on creatine kinases (McLeish and Kenyon, 2005). It is evident from the data of all isozymes (A thru D) that since the $K_d$ of the enzymes is lower for ADP, the enzyme affords faster utilization rate of ADP (i.e. synthesis rate of ATP). Therefore, if nature intended Complex V to be a physiological ATPsynthase, Complex V's affinity for ADP should have been higher (compared to ATP). Given the physiological ambiance, Complex V cannot deterministically prefer to synthesize ATP, when it is ordained to hydrolyze ATP. (For an extended discussion, please refer Item 2, Supplementary Information.)

*Table 1: Data on creatine kinases taken from Tables 1 and 2 of the review by McLeish and Kenyon, 2005.*

| Reaction | Constants | Isozymes | | | |
|---|---|---|---|---|---|
| | | *A* | *B* | *C* | *D* |
| **ATP hydrolys.** | $k_{cat}$ (x$10^3$ min$^{-1}$) | 9.2 | 12.9 | 3.1 | 4.5 |
| | $K_d$MgATP (mM) | 1.2 | 0.99 | 4.04 | 4.06 |
| | $k_{cat}/K_M$ MgATP (x$10^3$ mM$^{-1}$.min$^{-1}$) | 10 | 16 | 28 | 6.6 |
| **ATP synthes.** | $k_{cat}$ (x$10^3$ min$^{-1}$) | 29 | 21 | 4.7 | 5.4 |
| | $K_d$MgADP (mM) | 0.07 | 0.02 | 0.22 | 0.38 |
| | $k_{cat}/K_M$ MgADP (x$10^4$ mM$^{-1}$.min$^{-1}$) | 110 | 53 | 3.6 | 3.6 |

*Analysis of the prevailing ETC model:*

Since the pseudo-first order rate of substrate/oxygen utilization and ATP synthesis approaches ~$10^3$ s$^{-1}$, it is imperative that the rate of well-coupled ET (say, for one 4-electron relay across a set of Complex I – CoQ – Complex III – Cyt. *c* – Complex IV to reduce a molecule of oxygen) must approximate this value. This supposition is supported by literature (Orii, 1988). An elaborate analysis of each component and the concepts involved are discussed in Item 3, Supplementary Information. The following is a summation of the analysis.

The circuitry components (Complexes I through IV and Cyt. *c*; all added up) are in the range of $10^{11}$ - $10^{12}$ cm$^{-2}$ on the inner mitochondrial membrane (Gupte et al., 1984; Schwerzmann et al., 1986). In terms of volume, let's say that this would translate to roughly several decades' µM concentration of each protein complexes, which is not a conservative estimate, by any means. Now, let's take the ETC's multi-molecular sequential reaction scheme involving at least 15 collisions/interactions of 14 participants (of which 7 are distinct; NADH-**2**-Complex I-**2**-CoQ-**2**-Complex III-**4**-Cyt. c-**4**-Complex IV-**1**-O$_2$), each being assumed at a high concentration range of $10^{-4}$ M. Let's assume that even a single collision leads to a high-affinity binding and ET. Protons are also involved in each one of the 15 steps and they are available only at $10^{-7}$ M (and this is only in the bulk phase; in the trans-membrane space, it would be several orders lower). Let's forget the poor mobility of the bulky species involved and interfacial partitioning issues, and graciously assume a second order diffusion limitation regime of $10^8$ (for proteins) – $10^9$ (for protons) M$^{-1}$s$^{-1}$. A single step involving $10^{-4}$ M protein would maximally give a pseudo-first order rate of $10^4$ s$^{-1}$ and a single step involving $10^{-7}$ M protons would give a pseudo-first order rate of $10^3$ to $10^2$ s$^{-1}$. Now, in each one of the 15 steps, these two processes are involved and let us not forget that the ETC solicits that these steps are in an ordered sequence. It makes perfect quantitative logic to argue that 15 sequential steps (each with a limiting frequency of ~$10^3$ to $10^2$ hertz) cannot work hyper-concertedly to give an overall frequency exceeding $10^3$ hertz (the experimentally observed ET rates in mOxPhos). This statement becomes even more relevant considering that some reactions in the ETC scheme are two-electron or four-electron transfer steps (NADH/Succinate to Complex I/II, Complex I/II to CoQ, CoQ to Complex III, Complex IV to $O_2$), which would be considerably slower than the one-electron process. The crucial point is- anytime a bound species dissociates or does not collect the full quota of two or four electrons, the "circuit" is broken. Now, Table 2 captures the overall ETC. As per the prevailing ideas, the reduction of one molecule of oxygen at Complex IV (by a total of four electrons derived from a molecule each of NADH and succinate) minimally solicits the synchronous and tandem working (or continuous linking) of ~70 proteins/small molecules present on/across the phospholipid membrane. This ETC solicits that >24 redox active participants (the number of one- or two- electron redox active species within the purported ETC) must make >54 electron transfers (in batches of one or two electrons) across a collective path of >600 Å (the minimal conservative distance that 4 electrons must travel from NADH/succinate to $O_2$) within the protein networks alone. If we start with NADH as the sole reductant and include a minimal distance that CoQ and Cyt. *c* would have to commute within the inner membrane and inter-membrane space respectively (and also factor in the distance for CoQ to recycle), we must accept that each one of the electrons must undertake a journey of $>10^3$ Å across a predominantly low dielectrics path. Such a hyper-concerted proposal has little viability (with respect to probability, thermodynamics and kinetics) and the practical magnitude of distance and number of interactions should strengthen the inference. This statement is further consolidated when one takes into account the barriers provided by the existence of several unfavourable potential gradients and many instances where the two adjacent redox centers are > 12 Å away. By a conservative estimate (Table 2), more than half of the total of 54 steps (both inter and intra molecular transfers) could be deemed non-viable with respect to the overall reaction time scales. So, it is logical to infer that this purported elaborate ETC involving ~54 transfers and ~$10^3$ Å (for reducing a molecule of water) cannot be completed in sub-millisecond time-scales.

*Table 2: A comprehensive account of the prevailing ETC scheme for the reduction of one molecule of oxygen (by the circuit of Complex I – Complex II – 2 Complex III – Complex IV).*

| Element; protons pumped$ | Participants | Steps (2e+1e) | Overall distance (highest) (Å) | Distance per electron (Å) | Overall gradient [start (low, high) end] (mV) | Unfavorable steps | Non-"route" redox centers |
|---|---|---|---|---|---|---|---|
| **Comp I; 4** | 10 | 1 + 16 | 214 (16.9, 14) | 108 | -320 (-480, -150) +113 | 14 | 2 FeS (N1a = -233 & N7 = -314) |
| **Comp II; 0** | 6 | 1 + 8 | 105 (16, 11.9) | 56 | -31 (-260, +60) +113 | 4 | 1 heme (-185) |
| **Comp III; 4** | 6 | 0 + 12 (including 6 for CoQ recycle) | [170 (34 to 20) (including 100 for CoQ recycle)] x 2 | 35 (not including >50 for CoQ recycle) | +113 (-90, +300 ) +254 | 6 | Nil |
| **Comp IV; 4** | 6 | 0 + 16 | 120 (16) | 30 | +254 (+240, +320) +820 | 8 | Nil |
| **Overall; 12** | 24 | 54 | >750 (at least ten transactions are above 12) | ~230* | ~ -400 to ~ +800 (nine transactions unfavorable) | 32 | 3 |

*\* This is a highly conservative estimate by any means. The value for the conservative distance of a single electron travel within "the highly efficient" supercomplex (formed by Complex I – Complex III – Complex IV) would be minimally ~350 to 400 Å [[as can be seen in Figure 1c, in the review by Enriquez and Lenaz (Enriquez and Lenaz, 2014) or Figure 7b of the review by Kuhlbrandt (Kühlbrandt, 2015). So, a multi-disrupted four-electron travel even in this "optimized" but "un-insulated discontinuously wired" system would total a distance ~ 1500 Å. Distances are given in centre-centre and edge-edge. $ Value given is per two-electrons' passage through the Complex.*

## Discussion

***Fitting the "structure-function perspective" of the current findings with respect to known facts:***

Figure 7 depicts the RCPE-expected versus existent structural realities of Complex I, the most significant protein of the respiratory assembly.

*Figure 7: Top: A schematic representation of the ETC route and proton pumping sites within Complex I. Complex I has 9 Fe-S clusters and a flavin embedded within the matrix-ward projection. Of these, two are non-reducible and 2 fall outside the purported ET route. The proton-pumping sites are distant and disconnected from the electron transfer routes. Bottom: The expected structure of a sample of respiratory complex that could have represented the RCPE system. The prevailing ETC seeks a smaller matrix-ward projection (with a direct transfer of two electrons from the source to the membrane portion, without intervening solvent-accessible cavities), the burial of redox centers within TM region and specific relay of membrane-soluble redox relay agents. Then, such a series could have served as a kind of electron relay which could aid interspersed proton-pumping along the ET route.*

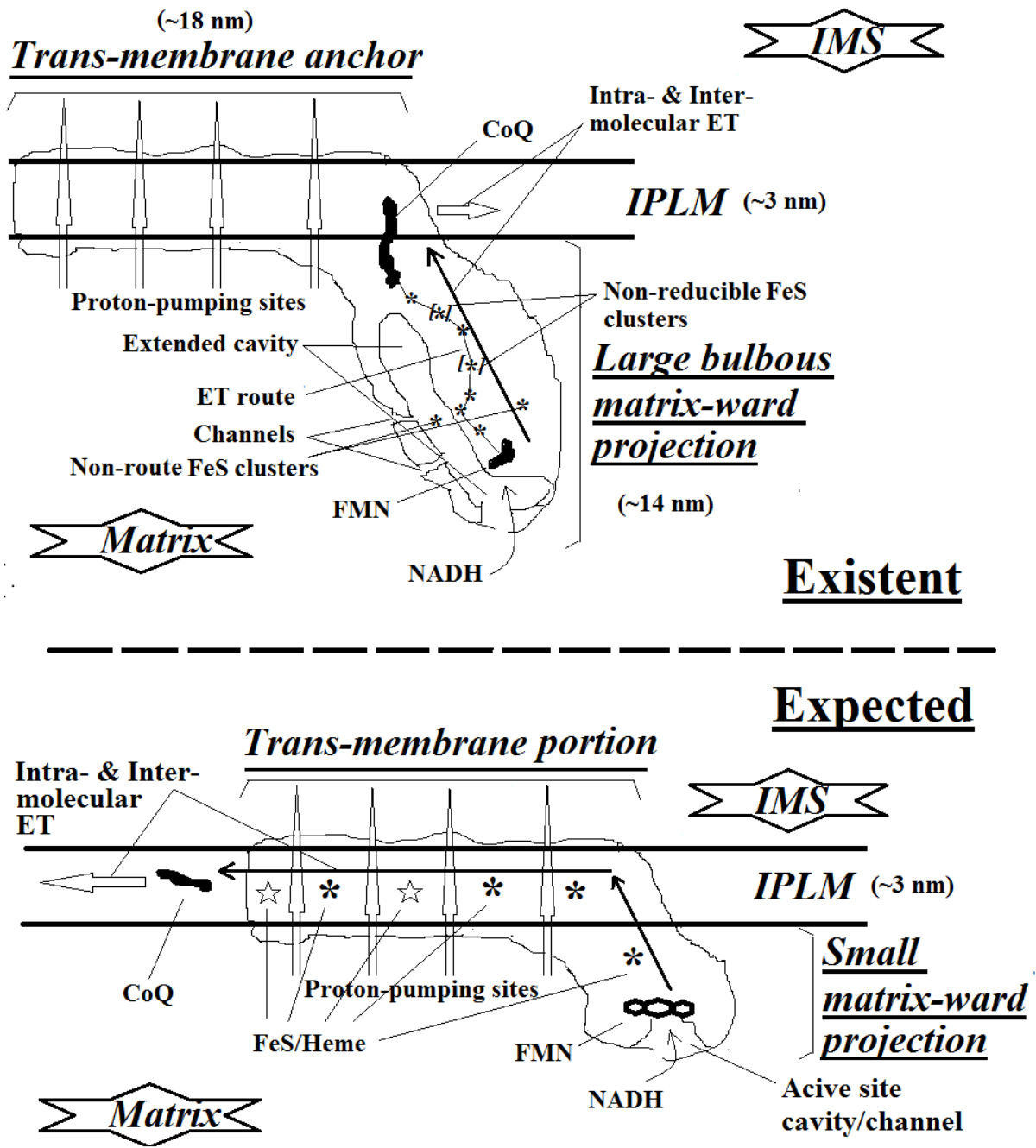

If the RCPE scheme was operative, one would imagine a series of redox centres within Complex I's horizontal foot, which could pump a proton across the membrane as the electron passed through it. But this is not the case as the protracted trans-membrane "foot" has no redox-centers. If Complex I's matrix-ward arm only served to conduct electrons from NADH to CoQ (as per the RCPE view), the multitude of redox centers, and the huge cavity with solvent connectivity would be highly disadvantageous, leading to electron leakage and production of DROS. It is quite a far-fetched mechanistic scheme to imagine that the transfer of two electrons from flavin to CoQ (an event known to occur within microsecond scales) would trigger the pumping of four protons across the "totally disconnected foot" of Complex I (an event which would occur within millisecond scales). Per the new oxygen-centric mechanism, Complex I is the major DROS generator/moderator within the system and the identification of multiple ADP-binding sites and orientation of channels justify the roles attributed to the same.

Since Complex II has only one ADP binding site and the DROS production ability is much lesser compared to Complex I (in terms of structural features, redox potentials and proton adequacy within $FADH_2$), ATP production would be lower via this protein's flavin-activation of oxygen.

RCPE hypothesis only attributed the role of Q-cycle to Complex III and the presence of a bulbous matrix-ward protrusion remained a long-time conundrum. The Q-Cycle seeks that three different molecules need to be concomitantly bound at three distinct loci on Complex III, and electrons must also flow from $CoQH_2$ back to CoQ. Such a scheme would have low proability and little drives. Since heme $b_H$ is accessible to "solvent", it is highly unlikely that such an elaborate scheme could be operative.

Instead, Complex III's structure would enable it to recycle (via $O_2$/DROS) and reset a one-electron paradigm in matrix from the electrons "lost" to CoQ. Further, with the identification of multiple ADP binding sites within the matrix-ward "bulb" of Complex III, the murburn explanation can more effectively justify this protein's role in mOxPhos.

Complex IV would be a major DROS generator/modulator (owing to the channel-accessible pentaligated heme and copper centres) and its location within the respirasome is adjoining a few ADP binding sites.

The finding that Complexes I through IV have ADP binding sites and DROS-channels is reminiscent of the explanation for cytochrome P450 activity in the mXM system, particularly the prolific reactivity of CYP3A4 (Manoj et al., 2016c; Venkatachalam et al., 2016). Overall, the respirasome super-complex makes more "structure-function" purpose if we assume the operational relevance of DROS. The current study has unveiled two high-affinity ADP binding sites adjoining the trans-membrane foot region of Complex I, which connects it with Complexes III and IV in the respirasome structure. Also, the Complex-III ward opening of the channels of Complex I's active site cavity is yet another structural feature that enables effective DROS-utilization. We can now understand that the individual components and the collective respirasome structure are evolution's way of continued optimizing of DROS-aided "decentralized" ATP synthesis via a one-electron paradigm. The supplemental ADP binding sites "re-discovered" in Complex V (originally reported thirty five years ago (Daniel and Zippora, 1984)) could also serve the newly allocated steady-state role of this protein.

Now, we shall discuss murburn mechanism and place our observations/deductions within the holistic context (with respect to literature).

*The case for an oxygen-centred reaction mechanism:*

The RCPE hypothesis does not give oxygen its "due merit and importance". Oxygen, a highly mobile/active species, must stay tethered at Complex IV, bidding time and resources to become water (after eight events transpire, wherein four protons and electrons are added). Such a process would not be hi-fidel, spontaneous, and fast. Under normal/physiological conditions, oxygen is available at $10^1$ to $10^2$ μM concentration in the aqueous cytoplasm and is at least 4-5 folds more soluble in lipids. Why should such a freely diffusible small molecule stay stuck to the reaction center of Complex IV alone? How can a single protein's binding mechanism evolve for holding on to the different species that would be formed? FTIR/Raman Spectroscopy experiments with several proteins' metal centers indicate that even the tight-coordinating diatomic ligands get displaced very easily, at low energy microcosms of even -100 °C. Then, if the heme $a_3$-$Cu_B$ center did evolve to perfection for the binding role, why does it get corrupted by a variety of ligands like CO, azide, cyanide, etc.? The lipid bilayer's permeability values of protons/hydroxide ions, water molecule and gaseous oxygen are in the range of $\sim 10^2$, $\sim 10^4$ and $>10^7$ nm s$^{-1}$ respectively (Milo and Phillips, 2015). It is known that the diffusion coefficient of protons/hydroxide ions, water and oxygen in aqueous systems are in the range of $10^9 - 10^{10}$ nm$^2$ s$^{-1}$. Therefore, the small dimensions, amphipathic nature and high motility of oxygen-centered entities (as exemplified by singlet or triplet oxygen, superoxide or hydroxyl radicals, hydroxide ion, etc.) could easily afford them a linear dimensional coverage radius of $\sim 10^1$ to $10^3$ Å/μs in the phospholipid interface, reaching out even into some occluded redox centers of proteins. Since effective electron transfer phenomena observed in the biological systems occur across a distance of ~10 Angstroms in micro- to milli- second time scales, oxygen-based small entities would be "practically everywhere at all time points" (with respect to the spatio-temporal considerations relevant to the reaction realm). There are more than two dozen known one-electron redox-active moieties and molecules (nucleotides, hemes, flavins, copper centers, organics, etc.) within the mitochondrial ET system. In physiological conditions, NADH/superoxide combine can "reduce and recycle" metal centers within the redox window of -320 mV to +460 mV (Pierre and Fontecave, 1999). It would be improbable that the highly mobile/reactive oxygen/ROS would not interact with these centers, particularly under "electron surplus" (reductive) conditions. There could be no ordering or deterministic scheme that could outdo this probabilistic predicament. Therefore, "nothing" within/across the phospholipid membrane could circumvent oxygen/DROS from shunting the proposed ETC circuitries, particularly given the structural aspects unveiled in this manuscript. In a suitable reductive ambiance, oxygen can receive one, two, three or four electrons (or hydrogen atoms), to respectively form superoxide, peroxide, hydroxyl radical + hydroxide ion and water (two molecules) (Wood, 1988). Spin conversion of oxygen in such metallo-flavin laden system is also facile. It is known that flavins can spontaneously activate molecular oxygen to give superoxide, Fe centers can produce hydroxyl radicals and Cu is an efficient singlet oxygen generator. No one will contest the fact that there is ample evidence for the formation of copious amounts such DROS by all ET complexes (I through IV) as demonstrated *in vitro*/*in situ*/*in vivo* (Bleier and Dröse, 2013; Dröse, 2013; Grivennikova and Vinogradov, 2006; Ksenzenko et al., 1992). Besides, there exists no real physical contiguity (wiring) in the ETC because CoQ and Cyt. *c* are freely soluble species within the electronic circuitry, both of which are functionally and spatially demarcated. CoQ is purportedly a two-electron relay agent within the inner phospholipid membrane whereas Cyt. *c* is a one-electron relay agent solubilised within the inter-membrane space of the mitochondrion. These predicaments dictate that the so-called ETC circuitry would be slow, easily broken, and it cannot prevent DROS formation. Therefore, it is now worthy to indulge the proposal that the mitochondrial system could have evolved to utilize the reactivity of DROS.

*Addressing the evidences for and against RCPE hypothesis:*

Before going on to the details of murburn explanation, it is opportune to address the experimental observations which were taken to support the chemiosmosis or RCPE view of mOxPhos and summate the case against RCPE hypothesis.

*How is the "indirect demonstration of proton pumps" explained? How are the experiments that vouched for ATP synthesis with an initialized "proton±ionic trans-membrane gradient" (in mitochondria/chloroplasts and reductionist models) be explained?*

(i) When an anaerobic mitochondrial suspension with succinate+ADP+Pi was given oxygen, the bulk extra-mitochondrial pH was found to drop suddenly within seconds and return to the original value within minutes. With a forthright deduction, this experimental outcome cannot be taken to support chemiosmosis. In fact, it should counter the "closed and coordinated system" perspective that Mitchell's postulates seek. If the "inner-membrane's proton pumps" worked, then the outer membrane is unable to contain the "surplus protons pumped out". As a result of this predicament trans-membrane potential cannot ever develop in a physiological setup. Further, if the internal membrane takes the time-scale of minutes to "spontaneously take the protons back in", then the physiological ATP synthesis cannot be explained with a spontaneous inward movement of protons via Complex V. The arguments presented above can be restated with another perspective- dynamics that occur in seconds-minutes scales within an "initialized system" cannot be given as evidence for a "physiological steady-state" event occurring in sub-millisecond timescales. A new explanation for the above-mentioned phenomena could be as follows. Isolated mitochondrial systems would have lowered or depleted ATP/NADH and lowered levels of Krebs' cycle metabolites. Succinate is both a Krebs' cycle metabolite and a fuel of mOxPhos. Among the Krebs' cycle intermediates, it is the dicarboxylic acid with the highest $pK_a$ values (4.0 and 5.24 respectively for the two acidic moieties). Input of oxygen into the milieu serves to replenish ATP levels and mitochondrial-metabolite transport systems also. This could lead to production of metabolites (with lower $pK_a$) leaching out and more ADP and Pi going in. Thus, these events are more likely to explain the decrease in extra-mitochondrial pH. Also, when dyes/indicators report that the matrix becomes alkaline, it might just imply a production of hydroxide ions in the inside (owing to the hydrogen-atom deficient nature of NADH), and the outcome need not be owing to "proton pumping activities".

(ii) Jagendorf's *in vitro* experiments showed ATP synthesis when a chloroplast suspension pre-equilibrated at alkaline pH was exposed to an acidic buffer. This outcome can be explained by the simple consideration that at higher pH, a closed membranous system with predominantly ADP+Pi would need protons for the synthesis of an ester bond, because the reactants' $pK_a$ values are near neutral pH. The phosphorylation reaction in this scenario is represented by (Ad = adenosine)-

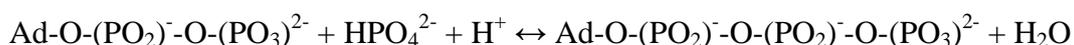

$$\text{Ad-O-(PO}_2\text{)}^-\text{-O-(PO}_3\text{)}^{2-} + \text{HPO}_4^{2-} + \text{H}^+ \leftrightarrow \text{Ad-O-(PO}_2\text{)}^-\text{-O-(PO}_2\text{)}^-\text{-O-(PO}_3\text{)}^{2-} + \text{H}_2\text{O}$$

Under these conditions, Complex V may work as an equilibrium driven "ATP synthase". A comparative analogy can be cited from lipase-catalyzed reactions. With normal aqueous micelles, lipolysis is favored whereas in reverse micelles with low water content, esterification occurs (Han et al., 1987; Hayes and Gulari, 1990). While lipase esterification is a simple reaction that can go both forward and reverse, "the interfacial machine of Complex V" cannot work in reverse under physiological conditions owing to- (i) the high affinity of Complex V to ATP & (ii) the synthase cycle being a multi-substrate reaction that requires

ADP, Pi and protons. The physiological reaction system for mOxPhos is faster and it occurs at high ATP concentrations and a range of pH values (and protons are at a premium within the mitochondria). The fact that atractyloside and bongkrekate inhibit ATP synthesis clearly points out that ATP-(ADP+Pi) equilibriums govern overall physiological catalysis in the mitochondria. Further, in a proton restricted environment, the thermodynamic drive for the redox reaction: $2NAD(P)H + O_2 \rightarrow 2NAD(P)^+ + 2OH^-$ goes several folds higher if external protons are available for water (O-H bond) formation. So, the enhanced ATP formation with a proton gradient in an "initialized system" can be theoretically explained by multiple rationales. The most important aspect to note is that a proton gradient cannot be generated and harnessed by mitochondria in "steady-state" (Manoj, 2017). Therefore, the above experiment has little relevance to physiological realms.

(iii) Racker's experiment had shown that a vesicular system with rhodopsin and Complex V gave ATP synthesis upon exposure to light. The first paragraph (second line) states that the membrane of *Halobacterium* incorporates only rhodopsin as protein. Subsequent research (some published from the same group) had shown that membrane fractions of such bacteria include other proteins with heme-flavin systems and that such systems also showed signature DROS-mediated non-specific phosphorylations and methylations of several moieties (Blaurock et al., 1976; Schimz, 1981; Spudich and Stoeckenius, 1980a; Spudich and Stoeckenius, 1980b). Also, there was no evidence that ATP synthesis resulted out of a proton pump type of activity of rhodopsin-embedded membrane preparations. The lysine-conjugated Schiff's base of this protein has a highly alkaline $pK_a$ (>13) (Govindjee et al., 1994) and it is unlikely that such a simple functional group can serve as a membrane-embedded proton pump. On the other hand, such systems are well-known to generate DROS (Chen et al., 2012), which is expected of photo-induced transformations in retinal (Aboltin et al., 2013).

Recently, vesicles incorporating oxidase (a known DROS generator) and Complex V were also shown to support ATP synthesis (Otrin et al., 2017). Therefore, the demonstration of ATP synthesis by "pure" Complex V + rhodopsin vesicles can be explained with DROS chemistry.

(iv) Employing valinomycin-$K^+$ with mitochondrial system for the demonstration of a chemico-protonic gradient powering ATP synthesis or estimation of "liberated protons" is debatable. This could also negate a fundamental of Mitchell's postulates which requires an intact inner impermeable mitochondrial membrane. It should be noted that the valinomycin-induced exchange of internal protons with external potassium ions remains more of an unexplained phenomenon, even now. For example- calcium uptake by mitochondria increased with metabolic inhibitors of Complex I, III & V; and also with valinomycin. But addition of high amounts of potassium ions lowered accumulated internal calcium. Further, it was found that valinomycin did not just facilitate $K^+/H^+$ ion replacement & diffusion-equilibration across the membrane, but the former's concentration determined the equilibrium position of the $K^+$ ion in/out distribution. It is pertinent to peruse Gilbert Ling's works/writings in this regard for some interesting experimental observations and key conceptual ideas on cellular homeostasis (Ling, 1981). Using monactin and valinomycin (in conjunction with added potassium ions), it was shown that the diffusion barriers offered by the inner mitochondrial membrane (as perceived) is not owing to a continuum of phospholipid layer. It is known that mitochondria readily exchanges cations (like sodium, potassium or calcium), has significant permeability features with respect to anions and also houses lots of aquaporin (Calamita et al., 2005). Besides the fact that the inner membrane has almost ~80% proteins, there is no direct evidence for "special features" or proton pumping nature of the inner mitochondrial membrane. Hence, it can be argued that all types of ionic species would be subjected to several intricate networks of equilibriums within a relatively closed system. The charge/power that can result out of ~50 nM concentration of protons (<10 protons per mitochondrion) would be insignificant with respect to the other ionic species that exist in direct equilibrium within. So, introduction of "valinomycin-$K^+$"probe is not only antithetic to the hypothesis being ratified, application of the same does not allow us to trace the "cause-consequence" correlation.

***TMP has been observed to build up and correlate to mitochondrial ATP synthesis. Isn't that evidence for Mitchell's and Boyer's hypotheses?***

(v) Mitchell postulated a steady-state trans-membrane potential operating between the matrix and inter-membrane space of a physiologically active mitochondrion. This has not yet been demonstrated and cannot be verified in the foreseeable future because of "technical difficulty" of finding a fine probe that could be introduced into the inter-membrane space. The TMP observed, which was taken as support for Mitchell's hypothesis (Junge et al., 1968) is between bulk aqueous phase and mitochondrial matrix. Most importantly, there is no solid theoretical or structural or quantitative justification of how a trans-membrane potential could be practically transduced to give ATP synthesis in physiological states. Elasticity-based or surface energy-based (Junge's/Warschel's/Nath's works) simulations/calculations merely build on the Mitchell-Boyer "assumption" that Complex V is the ATP synthetic agent in mitochondrion that can tap a trans-membrane potential. The man-made dynamos/motors that harvest potentials via a cyclic modality use the principle of electromagnetic induction. This would need precisely arranged components (including ferromagnetic parts) arranged via an intelligent and directional agent, and the whole setup must work in a synchronously, in a staggered modularity. Complex V does not have any of the needed "intelligence" or structural or compositional attributes. On the other hand, it is common knowledge that the faster a motor or generator runs the greater noise or smoke it may produce. The noise or smoke could be analogous to generation of TMP in mitochondria. Inferring that the TMP is the causative force leading to ATP synthesis could be erroneous. In other words- higher TMP observed with ATP synthesis is merely coincidental, and not consequential.

***Doesn't the requirement for intact mitochondrial membrane support chemiosmosis? Doesn't chemiosmosis explain the effects of ionophores, uncouplers, etc.?***

(vi) The "non-synchronized action" of the purported proton pumps in itself should serve as an uncoupling agent, if RCPE mechanism held true (Manoj, 2017). Uncoupling molecules with similar structures (i.e. disubstituted phenolics) were also found to inhibit mXM system, and the mXM system does not necessitate any proton pump or intact spherical membranes (Manoj et al., 2016c; Parashar et al., 2014a; Parashar et al., 2018). Thus, an uncoupling effect is observed probably due to interfacial modulation of the essential DROS dynamically generated in the system. An uncoupler like dinitrophenol would have low mobility across the highly "impermeable" membrane, particularly owing to the charge on its nitro groups. One wonders how or why they should keep dissipating an assumed gradient across the inner membrane. If it did, why is the intact mitochondrial ETC (oxygen uptake) dependent on ADP & Pi, when it is known that uncouplers/ionophores can delink electron transfer from ATP synthesis? Seen in another perspective- mitochondrial fragments could carry out electron transfers, but not ATP synthesis. How does RCPE hypothesis come to terms with the fact that intact mitochondrial systems need ADP & Pi, when the ETC does not need them at any stage? It is a puzzle how/why the incorporation of ionophores to determine H:P ratios does not violate Mitchell's own postulates. As pointed out earlier in this writing and elsewhere, such experimental procedures and inferences must be doubted (Ling, 1981; Nath, 2010a, b; Slater, 1987). The requirement for a closed lipid membrane system is essential to make a low-water/proton environment that could aid the equilibrium requisites and effective reaction zone enabling DROS-mediated catalysis.

*How is the recent "single molecule" experiment with Complex V explained?*

(vii) In recent times, studies at single-molecule level have "shown" direct rotation of Complex V (Rondelez et al., 2005; Sielaff and Börsch, 2013; Tanigawara et al., 2012; Volkán-Kacsó and Marcus, 2016). Using a His-tag tethering of the $F_1$ module to a slide, the demonstration of rotation of a tagged actin filament attached to the γ shaft cannot be deemed functionally analogous to the proposed physiological functioning. Please refer Results section and Item 2 of the Supplementary Information for pertinent arguments regarding the context. Since we know that the electron transfer from NADH to oxygen is physiologically linked to ATP synthesis, the most relevant question is- When the intact mitochondria do not show oxygen consumption without ADP + Pi but disrupted mitochondria consume oxygen without being presented with ADP + Pi, doesn't that imply that the physiological mOxPhos routine is quite a different reaction system?

*<u>So, what is the "case summary" against RCPE hypothesis?</u>*

*<u>Against ETC</u>*

\* highly sequential scheme that is supposed to operate without a thermodynamic "push or pull"

\* a "vitally deterministic and fastidious" model that repeatedly solicits multi-molecular complexations (as exemplified by CoQ cycle)

\* role allotted for oxygen (serving as a terminal electron acceptor, staying wedded to Complex IV) is rather insignificant

* the actual kinetics of physiological electron transfer is too high and anoxic electron transfer rates are too low in experimental systems (that is- outer sphere model of electron transfers cannot explain overall water formation rates) (Orii, 1988)

* order of arrangement of redox centers of varying potentials and distances between the redox centers in/across protein Complexes goes against a viable ETC

* there exists no logic as to why so many redox centers and proteins are required in the overall ETC

* why should there be "non-route" redox centers (as exemplified in Complexes I & II)?

* why are some redox centers within Complex I not reduced (as exemplified by N5 and N6b) even in the most favorable conditions?

* natural mobility/reactivity of oxygen and other organic or inorganic molecules or protein complexes within the system must be overlooked for ETC to function (molecules need to function by "over-riding their natural dispositions")

* presence of DROS in actively respiring mitochondria and in reconstituted systems (with individual Complexes) affords neither marginal utility nor esthetic appeal to ETC

* how several synthetic dyes and redox active molecules can put in and receive electrons from ETC (and why oxygen messes electron transport in anoxic bacteria)

* inability to explain electron transfer rate variations (and electron leaks) in different "metabolic states", particularly states 2 and 3

* binding to Complex IV cannot explain the toxicity of low amounts of cyanide

* why both low and high amounts of oxygen lead to "oxidative stress" and ROS-induced damages

* $K_d \gg K_M$ conundrum of Complex IV-oxygen interaction (Orii, 1988)

*Against Proton Pumps*

* low proton availability in mitochondria [<10 protons per mitochondrion; when >100000 are needed! (Manoj 2017)] chokes the proton pump proposal

* little direct or unambiguous structural and functional evidence for proton pumps

*Against ETC-Proton pump combine*

* there is no one-proton : one-electron correlation for the purported proton pumps

* the electron transfer routes are located away from purported trans-membrane "proton pumps"

* the electron transfers are usually in microsecond timescales whereas trans-membrane proton transfers are envisioned in millisecond timescales

*Against Chemiosmosis*

* chemiosmosis is an absolutely untenable proposal with respect to thermodynamics

* ATP synthesis can be obtained in mitochondria incubated at pH higher than neutral values

* ATP synthesis is noted in the absence of proton gradients and also at low trans-membrane potential

* if buffering is operational in matrix, chemiosmosis cannot work in steady state

* steady state perspective cannot be realized as it seeks little protons in the matrix for gradient build up but copious amounts of protons for proton pumps to function (i.e. the explanation seeks the impossible realization of two mutually exclusive options)

*Against ETC-Rotary ATP synthesis combine*

* how does blocking of Complex IV by cyanide inhibit ATP synthesis completely in systems with higher levels of oxygen?

*Against Proton Pump-Rotary ATP synthesis combine*

* how could proteins (proton pumps and ATPsynthases) change conformations in a steady state of constant unipolar trans-membrane potential (no temporal window for "native-excited-native" conformational changes)

*Against Rotary ATP synthesis*

* Cohn's observation of multiple $^{18}$O atoms' incorporation into ATP (Cohn, 1953)
* lack of evolutionary justification for how such a highly sophisticated rotary enzyme could be formed in the initial phases of life
* no thermodynamically compatible explanation of how Complex V could function as an enzyme that could aid ATP synthesis within the mitochondria (when its demonstrable function is that of an ATPase)
* why does the $F_1$ subunit have several order higher affinities for ATP compared to ADP?
* little rationale for how Complex V taps into a trans-membrane potential in steady state
* how could the $F_o$ subunit bind to a frail lipid membrane and rotate at the same time

*Against RCPE in toto*

* non-modular organization of mitochondria with no staggered synchronization scope therein for the development of a TMP based on matrix-to-outward pumping of protons
* relative distribution densities of mitochondrial respiratory complexes and components do not support RCPE
* how could the operational logic of RCPE (with gambits and "irreducible complexities") evolve from a minimal set of components

* unclear as to why Complex III would take two electrons from $CoQH_2$ and give it to two molecules of Cyt. *c* across the membrane as this appears not just "unnecessary", but also "counter-productive" with respect to the purpose of RCPE

* variable and non-integral stoichiometry goes against the definitive/ordered mechanism

* maverick dose responses proffered by various molecules does not agree with specific binding-based reaction outcomes

* requirement of ADP + Pi in the intact mitochondrial system for oxygen uptake when disrupted systems do not need ADP + Pi for oxygen uptake

* proton-deficient NADH as the "evolutionarily chosen" molecule for the role of reductant

* heat generation by amphipathic uncouplers and uncoupling protein of brown adipose tissue

* why leaching of cytochrome *c* leads to cellular apoptosis and how synthetic vesicular systems could work without Cyt. *c* (when RCPE needs Cyt. *c* to be present in the inter-membrane space)

* the favorable roles of extensive amounts of constitutive anionic lipids like cardiolipin in mitochondrial membranes

* inhibitory roles of molecules like oligomycin and venturicidin (compared to valinomycin)

* failure to satiate a simple and facile chemical reaction logic for "coupling" and inability to meet biological structure-function correlations

* the efficiency of ATP synthesis by RCPE is too low (1.5 for succinate and 2.5 for NADH) to account for the higher values obtained experimentally

* the toxicity of cationic lipids (as seen in drug delivery efforts; (Aramaki et al., 2001)) to mitochondrial metabolism is not explained

* there seems to be little rationale for electron(s) transfers in pairs or as unit entities criss-crossing the membrane through respiratory complexes, and for physical transport from one

complex to another (via CoQ and Cyt. *c*) only to form water at the end of the exercise (please refer Figure C1, Item 3, Supplementary Information)

* kinetically inexplicable how hundreds of proteins/small-molecules/ions (found varying at $10^{-3}$ to $10^{-8}$ M) distributed across three phases (matrix, inner membrane, inter-membrane space) work together sequentially to make >50 electron transfers and channelize ~$10^4$ protons out/in the inner membrane via specific routes, to make $10^3$ molecules of water/ATP in a second

As seen from the points above, not only are the four components of the RCPE paradigm untenable independently, they are also incapable forming a cohesive whole. Since most hitherto presented "supportive evidences for RCPE" have also been discounted, it is now imperative to think beyond this paradigm.

*Introduction to murburn concept:*

Generally, redox proteins were thought to function physiologically quite akin to the other classes of enzymes, via an enzyme-substrate affinity-based binding at the active site. Further, transfer of electrons within or between proteins was also believed to occur solely via preferred routes "wired" within the components (for inter- and intra-molecular electron transfers). While probing redox enzyme mechanisms, my group's works showed that peroxidases/P450s'catalyses were inhibited/modulated by agents that affected the dynamics of diffusible and reactive oxygen species (DROS) (Andrew et al., 2011; Gade et al., 2012; Parashar et al., 2014a; Parashar and Manoj, 2012; Parashar et al., 2014b). We had also demonstrated that the flavoenzyme cytochrome P450 reductase generates DROS from NAD(P)H (Manoj et al., 2010b) and a diverse array of heme-proteins could avail DROS to carry out physiologically viable enzymatic turnovers, without getting deactivated (Manoj et al., 2010a; Manoj and Hager, 2001; Manoj et al., 2016c; Parashar et al., 2018). Therefore, quite contrary to the prevailing perceptions, it was inferred that several electron transfers, catalyses and inhibitions observed in heme-flavin enzymatic systems may not obligatorily involve high-affinity "protein-protein and protein-small molecule" complex formations mediated via "active-sites" (Gideon et al., 2012; Manoj, 2006; Manoj et al., 2010a; Manoj et al., 2010b; Manoj et al., 2016a; Manoj and Hager, 2008; Manoj et al., 2016b; Manoj et al., 2016c; Manoj et al., 2016d; Parashar et al., 2014a; Parashar et al., 2018; Parashar et al., 2014b; Venkatachalam et al., 2016). Thus, DROS was shown to be a mild cum non-specific *in situ* agent for electron transfer and redox-catalytic metabolic routines. The entailing murburn process invoked upon a moderated "molecule-ion-radical" dynamic/stochastic interactive paradigm (Figure 8A) (Manoj, 2017; Manoj et al., 2016a; Manoj et al., 2016b; Manoj et al., 2016c; Manoj et al., 2016d; Parashar et al., 2018; Venkatachalam et al., 2016). The new understanding allowed us to explain unresolved aspects of microsomal xenobiotic metabolism (mXM) (Manoj et al., 2016c) and the conundrums of hormetic cum idiosyncratic physiological dose responses (Parashar et al., 2018). Although "chaotic at high densities", the diffusible species mediated one-electron reaction scheme is very reproducible, selective and specific at discretized space-time scales (Manoj, 2017). This idea is not very "radical", considering that enzymes like ribonucleotide reductase and systems with cyanocobalamine are accepted to work through a radical mechanism (wherein the substrate is held via high-affinity recognition, within a few Angstroms of the site of radical produced on the protein).

*Figure 8: Generalized oxygen-centric paradigm and the specific application to mOxPhos system: **A.** The redox enzyme molecule may possess a source of "d or/and π" electronic*

*systems that can be destabilized/delocalized. A confined zone of reaction is mandated failing which the radicals cannot be harnessed. The reason could be a "caging" effect or the probability that the radical could meet another one such and thereby collapse. The ions could relay the electronic charges or stabilize the intermediates. High affinity physical contact between the initial donor (enzyme/substrate) and final acceptor (substrate/enzyme) is not obligatory.* **B.** *The differences between the electron transfers of RCPE paradigm and the new reaction scheme for mOxPhos are depicted. The ETC perspective seeks a primarily two-electron flow through a "pipeline" of Complexes I through IV, and oxygen stays committed to Complex IV. The new understanding directly connects electron transfer and ATP/water formation (via each one of the individual respiratory complexes). In the new perspective, oxygen does not stay "wedded" to Complex IV alone, but DROS species are involved at each individual component of the respiratory "battery", an assembly of membrane (super)complexes. The electron relay is faster with the formation of ATP and water, and this is by the virtue of a "thermodynamic pull". (For more mechanistic details, please refer text.)*

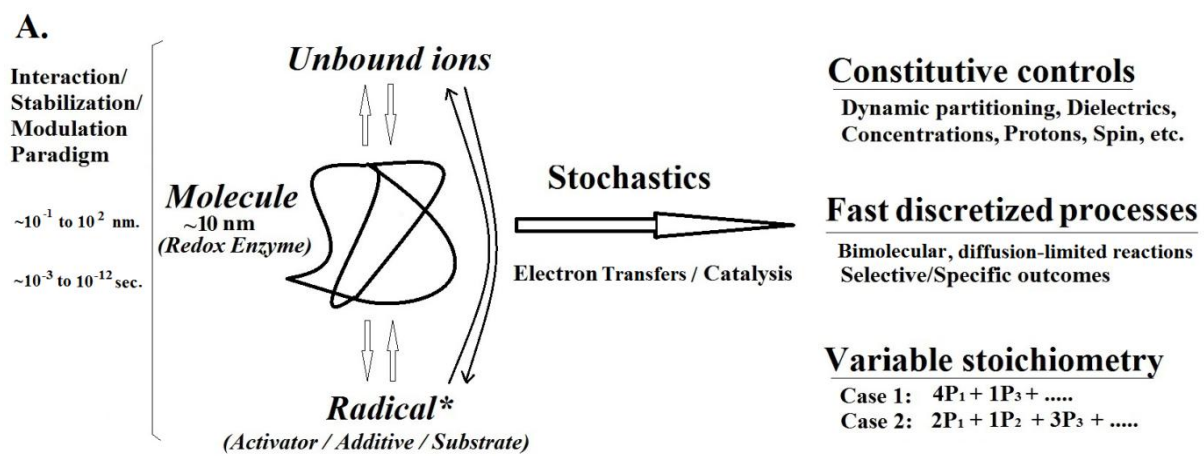

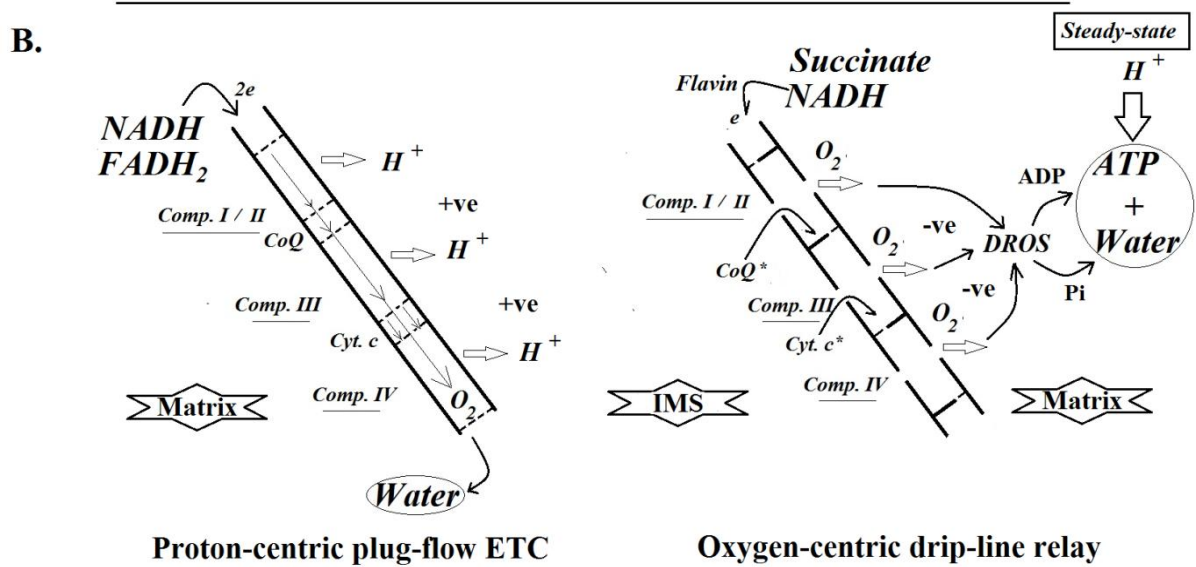

The mXM system is a multi-protein array embedded within the endoplasmic reticulum membranes of hepatocytes, with several hemeproteins- isozymes belonging to cytochrome

P450 (CYP) class and cytochrome $b_5$ (Cyt. $b_5$), and small amounts of a unique diflavoenzyme cytochrome P450 reductase (CPR). It was recently pointed out that the structural/distributional features of components and overall schema within the mXM and mOxPhos systems were markedly similar (Manoj, 2017). It is well-known that DROS production/detection has been positively correlated to $O_2$ and NADH concentrations (Murphy, 2009). Also, DROS production/detection goes up with ATP synthesis and build-up of TMP in the mitochondrial system (Nicholls, 2004). Therefore, it was reasonable to argue that the new mechanistic explanations could be relevant in the mOxPhos system too. That is- quite like the DROS stabilized/modulated in the mXM system was demonstrated as a key agent for CYP+CPR+Cyt. $b_5$ aided hydroxylations, the dynamically formed DROS in mOxPhos system could serve as the coupling/phosphorylating agents for *in situ* ATP synthesis. This novel "oxygen-centered and DROS-mediated chemical-coupling" perspective could also potentially explain why earlier researchers (1950s to 1970s) were unable to identify an "enzyme-linked high energy phosphorylating intermediate" (Manoj, 2017) within the mOxPhos machinery.

Usually, an enzyme-catalyzed reaction mechanism falls under the purview of enzyme-substrate binding-based paradigms such as- Fisher's "lock & key" concept, Koshland's "induced fit" concept, Michaelis-Menten kinetics, Eyring's transition-state logic, etc. Let's call such facets as "classical" elements. The new catalytic mechanism may invoke the classical elements above, but the reaction scheme also incorporates an "enzyme-free diffusible species". That is- an enzyme could produce or stabilize or modulate a diffusible species (like DROS), and such species could finally react with a transiently bound or freely diffusing final substrate present in the vicinity. Let's call this aspect of the new explanation a "maverick" element. Therefore, in the oxygen-centric murburn paradigm, an "uncertainty" exists in determining/predicting the precise schema of interactions. As a result, the overall reaction may show adherence to classical schemes at times (when the maverick step is not rate limiting or chaotic), and yet behave unpredictably at other instances (when the maverick step becomes rate limiting or gets chaotic). Such eventualities result because the diffusible intermediate could sponsor secondary reactions and the intermediates thereby formed could also go through alternate modalities of cycling ("uncoupling"), not leading to the "product of interest".

Till date, murburn concept has been floated as a mechanistic proposal for three systems and their summary is presented in Table 3. It is very interesting to note that all these systems require extraneous protons for product/water formation and they show a constitutive "redox or thermodynamic pull" operative within the system (Manoj, 2017; Manoj et al., 2016a; Manoj et al., 2016b; Manoj et al., 2016c; Manoj et al., 2016d). That is- a small wastage of redox equivalents is inherent in the system even when the "final substrate" is not present. When the final substrate is added, the electron sources are depleted at a much quicker pace for the "constructive" reaction. The overall essence is that some redox enzymes could establish a small region around themselves where there exists a finite probability of finding a diffusible radical. This radical could interact with various other ions/small molecules, and if there are suitable ways to "sink" the electrons (from a source) within immediate vicinity, effective electron or group transfers could be achieved in the milieu.

*Table 3: Comparison of heme-peroxidase chlorination, mXM hydroxylation and mOxPhos phosphorylating systems.*

| System | Cofactors | Ions, pH & ambiance | Electron source/ Substrate | Major product | Minimal equation |
|---|---|---|---|---|---|
| *Heme-haloperoxidase* | Fe-Heme ($Mn^{2+}$) [One protein with a single cofactor (Manganese is optional)] | $H^+$, $Cl^-$; acidic pH; Purely aqueous system | $H_2O_2$; primarily hydrophilic | Halogen atom transfer, and/or substrate oxidation. [$R^{*+}$, R-R, RCl, RCl-R'OH, R=O, etc.] | RH + $L^-$ + $H_2O_2$ → RL + $H_2O$ + $OH^-$ |
| *mXM* | Fe-Heme (P450, Cyt. $b_5$), Flavins (reductase) (Fe-S centers in auxiliary systems) [Three proteins and four cofactors] | $H^+$, $OH^-$; neutral pH; Interfacial system, uncoupling high without lipid embedding | NADPH + $O_2$; primarily hydrophobic | Oxygen atom insertion or hydroxyl moiety transfer. [ROH, R=O, etc.] | RH + NADPH + $O_2$ → ROH + $NADP^+$ + $OH^-$ |
| *mOxPhos* | Complexes with Fe-hemes, Flavins and Fe-S centers, quinones, etc. [Several tens of proteins and cofactors] | $H^+$, $OH^-$, $HPO_4^{2-}$, ($Mg^{2+}$); values around neutral pH; closed vesicles with membrane-laden catalysts | 2NADH + $O_2$; amphipathic R-$(PO_2)^-$-$O^-$ | Phosphate group transfer. R-$(PO_2)^-$-O-$(PO_3)^{2-}$. | *Discussed in text* |

## *The murburn proposal for mOxPhos:*

Figures 8B and 9 provide a simple generalized scheme for the new model and its application in the mOxPhos system. As seen, the new paradigm connotes a localized one-electron reaction scheme that involves oxygen-centered molecules/ions/radicals, diverse electron donors (reductants) and stabilizers/ moderators/ modulators (protein complexes). While the erstwhile "electron leaks" could be explained by the prevailing model of ETC (wherein DROS is formed and they react amongst themselves), the metabolic ATP synthesis electron transfers are better reasoned by the relay process, as shown in Figure 8B. Therefore, the two-electron process of the erstwhile scheme serves as local 2e sinks and electron retrieval/recycle measures of mOxPhos.

In the murburn mechanism for mOxPhos, TMP arises because negatively charged DROS accumulate within the mitochondria. For example- the primary reaction 2NADH + $O_2$ → $2NAD^+$ + $2OH^-$ may occur via a transient formation of superoxide ($O_2^{*-}$) radical anions, thereby explaining the experimentally observed TMP and pH shifts. Such a TMP cannot be directly harnessed by Complex V for the phosphorylation process. In the oxygen-centric mechanism, the transient radicals (DROS) formed in milieu mediate the phosphorylation reaction. Therefore, DROS have obligatorily constructive roles within mOxPhos. A sample outcome of a one-electron cascade (post the formation of FADH$_2$ by succinate dehydrogenase) in the mOxPhos system is given in the bottom inset of Figure 9.

---

*Figure 9: Overall schema of mOxPhos: The physiological reaction would involve the generation, stabilization and modulation of diverse DROS species, which may attack Pi/ADP, charging the same and this could subsequently lead to the formation of ADP-Pi adduct. The overall process could be helped by Mg$^{2+}$ ions. Complex V could serve to usher protons in (owing to the proton limitation in NADH and/or enhancing radical attack on phosphoryl groups by proton supplementation in its vicinities) for completing the last step of ATP synthesis and water formation. The majority of the ATP and water thus formed would go out through ATP-ADP porting system and aquaporins respectively, and the steady-state reaction would continue. Therefore, dynamic partitioning of electrons (by a one-electron scheme) away from the source (NADH) forms the "redox/thermodynamic pull", which keeps the reaction going at a "diffusion-limited pace". The erstwhile ETC would be functional, but only as a slow back-up measure to recycle the electrons lost at each phase.*

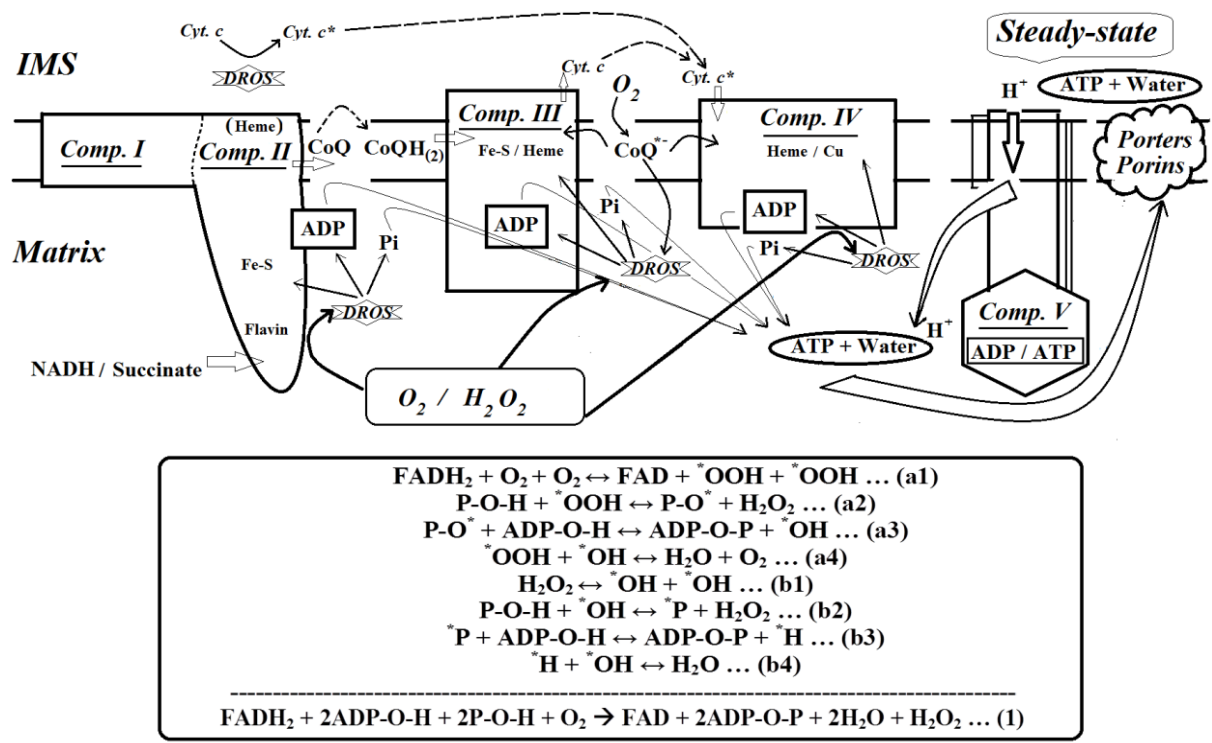

$$FADH_2 + O_2 + O_2 \leftrightarrow FAD + {}^*OOH + {}^*OOH \ldots (a1)$$
$$P\text{-}O\text{-}H + {}^*OOH \leftrightarrow P\text{-}O^* + H_2O_2 \ldots (a2)$$
$$P\text{-}O^* + ADP\text{-}O\text{-}H \leftrightarrow ADP\text{-}O\text{-}P + {}^*OH \ldots (a3)$$
$${}^*OOH + {}^*OH \leftrightarrow H_2O + O_2 \ldots (a4)$$
$$H_2O_2 \leftrightarrow {}^*OH + {}^*OH \ldots (b1)$$
$$P\text{-}O\text{-}H + {}^*OH \leftrightarrow {}^*P + H_2O_2 \ldots (b2)$$
$${}^*P + ADP\text{-}O\text{-}H \leftrightarrow ADP\text{-}O\text{-}P + {}^*H \ldots (b3)$$
$${}^*H + {}^*OH \leftrightarrow H_2O \ldots (b4)$$
---
$$FADH_2 + 2ADP\text{-}O\text{-}H + 2P\text{-}O\text{-}H + O_2 \rightarrow FAD + 2ADP\text{-}O\text{-}P + 2H_2O + H_2O_2 \ldots (1)$$

Herein, ADP-O-P is ATP made from ADP (≈ ADP-O-H) and inorganic phosphate, Pi (≈ P-O-H). Both Pi and ADP can be attacked, particularly if ADP is presented effectively to at the source of DROS. The redox centres within the mOxPhos assembly could also sponsor a scission of peroxide, as shown in (b1). Further, if the second electron from peroxide reacts with ADP/Pi and any hydrogen radical formed is stabilized/recycled via an oxidase function again, the yield of ATP would go higher. The scheme would go on, until the overall bond energetics' differential (products-substrates) stays viable. A statistical realm where the combination of several such reactions occurs in a single pot would give variable and non-integral stoichiometry. For example, in the above sample scheme, if one reaction got terminated at (a4) and the other went all the way until (b4), then we would have a yield of $[(1ATP / O_2) + (2ATP / O_2)] = 1.5$. If most reactions continued beyond (b4), we could also have ATP yields above 2.0 for $FADH_2$. The new scheme can potentially give >4 and >2.5 ATP per NADH and $FADH_2$ respectively, as opposed to the limiting values of 2.5 and 1.5 set by the proton-centric paradigm (Mitchell, 1975). Thus, the yield of mOxPhos is enhanced by combining one-electron "oxidase+peroxidase" function within the metabolic scheme. So, water formation is not via two independent two-electron reactions (like $FADH_2+O_2$ and ADP+Pi). Via DROS, both coupled ($ATP+H_2O$) and uncoupled reactions (heat+$H_2O$) result. In the uncoupled reaction, redox equivalents generated from fuel are lost because DROS react among themselves. In this regard, reaction like (b2) which leads to the production of "stable DROS" was inadvertently understood as an undesirable outcome within the milieu. The proposed bimolecular reactions with practically diffusion-limited rates are fast and efficient. Also, in all terminated reactions, the ratios of solutes (reactants : products) ranges from ~2:1 to ~4:1. Therefore, in steady-state, turgor/osmotic pressure (a colligative property) would enable water (the solvent and product, both) to spontaneously move out of mitochondria through the aquaporins (thereby serving as the thermodynamic "pull"/drive) and cations/protons could spontaneously or actively move in to neutralize the negative charges. Such a simple chemical coupling logic explains why ADP+Pi increases oxygen uptake (which cannot be explained by RCPE, because there is no direct chemical connectivity between NADH oxidation and ATP synthesis) and how toxins like atractyloside and bongkrekate inhibit ATP synthesis. Nature just tapped into the logic of colligative properties, facilitating steady-state dynamics by a turgor/osmotic movement of water from matrix to outside. Per the new perspective, Complex V is attributed a trans-membrane (matrix-ward) proton-delivery function (Manoj, 2017). This function could also serve as the "coupling factor in steady-state" because oxidation of NADH would be a proton deficient scheme, unlike the $FADH_2$ reaction shown in (a1). Therefore, the ATP-adducts formed through schemes such as the ones leading to equation (1) could need neutralization with protons, failing which the "bound ATP" may not detach (analogous to an "unpacked purchase" not being delivered at the cash counter). This explanation could fit the 6 ADP binding sites per $F_1$ subunit, the preponderance of Complex V and the high affinity reported for Complex V-ATP interaction. The electrophilic DROS intermediates could also "home in" better on the polar phosphates too. Electrons input from the relatively lesser efficient Complex II system would be primarily to enrich the $CoQH_2$ population within the membrane, so that the membrane system is "chilled" from oxidizing the lipid components therein. Thus, the new explanation (as shown in Figures 8b and 9) is favoured by Ockham's razor and agrees with evolutionary perspectives, mitochondrial architecture, membrane-embedded proteins' structure and distribution, abundance/mobility/reactivity of oxygen/DROS at the phospholipid interface, paucity of protons within membrane/matrix, reported kinetic isotope effects, etc. Now, mitochondria could have ATP hydrolysis and esterification happening side-by-side, with two different mechanisms- the one-electron synthesis and the two-electron hydrolysis.

Quite unlike the proton-centric RCPE hypothesis, the new oxygen-centric mechanistic model for mOxPhos is unordered, with neither sequential schemes nor multi-molecular complexations. The aesthetic perception that radical mediated reactions could be deleterious to life is "inconsequential" because DROS are an observable reality of cellular physiology and their roles fit the context's chemistry/physics. Further, discrete radicals need not be chaotic and do not attack the cellular machinery as long as other DROS (like peroxide) or more favorable "substrate" molecules are present in milieu. This statement has been validated in chloroperoxidase (Manoj, 2006) and mXM (Manoj et al., 2016c) systems earlier. If we assume that life originated from chaos, the stochastically regulated murburn scheme also serves the mandate of evolution. The essential "perception change" can be aided by the following analogy. When a cloth dipped in oil is set on fire, the fabric gets charred only after the oil is burnt. Similarly, the lipid cellular machinery is spared (in a statistical temporal sense) because relatively stable DROS (like peroxide) or more favorable "substrate" molecules are present in milieu for the radicals to get scavenged. Life evolved by exploiting the efficient stabilization/partitioning environment of metallo-/flavo- proteins embedded in phospholipid interface, which afforded the scope for harnessing the kinetically and energetically viable prospects of oxygen-centered radicals generated *in situ*. These could have well-aided the dynamic requisites of life at a molecular level and it is unlikely that evolutionary struggles could have sidelined such profitable prospects.

For a comparison of the operative logics of RCPE and murburn explanations, please refert Item 4, Supplementary Information. In the current study, the structural features of complexes that support murburn concept have been unveiled in the respiratory complexes. Further, more support for murburn concept is listed with respect to other structural aspects, evidence (collected from literature) for oxygen/DROS-centric reaction chemistry's relevance in physiological phosphorylations, the mechanism of inhibition by cyanide and explanations for miscellaneous inhibitions/uncoupling, solving other long-standing conundrums in the pertinent field, and stitching together evolutionary and holistic perspectives. (Key: # verified consequence, • argument supporting theoretical consistency, * potentially verifiable prediction/projection)

*Explaining structural and compositional realities:*

# The new model of mOxPhos would seek low levels of flavins and high amounts of hemes. As expected, the distribution of proteins shows low amounts of Complexes I & II (DROS generators) and high amount of Complex IV (DROS stabilizer/modulator) and Complex V (ATP binder/proton provider) (Gupte et al., 1984; Schwerzmann et al., 1986). Further, the relative concentrations of Cyt. *c* and Complex IV also align well with the new perceptions. Cyt. *c* and CoQ have been primarily allocated one-electron scavenging/stabilization and recycling roles within the IMS and IPLM. Their nature and location agrees with the roles allotted to them by the new paradigm.

• The organization of mitochondria with only cristae invaginations, two membranes and the membrane-proteins scattered therein agrees well with the new mechanism.

• Murburn concept explains why the Complexes I through IV have several tens of redox centers. Also, why some redox centers N1a and N5 in Complex IV are not in the "route" or not aligned with a perfectly/strictly increasing order of potentials is better explained as a deliberate ploy by evolution for aiding DROS formation.

• The mitochondrial membranes are seen as a hydrophobic interface that could potentially support and confine radical reactions. The erstwhile consideration required the inner mitochondrial membranes to be discriminatory and regulatory in structure and function. For all practical purposes, mitochondrial membranes don't show "super-deterministic" ability/features.

• The influence of membrane lipids like cardiolipin (Paradies et al., 2014) can be better explained with the new paradigm. Their high negative charge densities on their lipid-assembled periphery would keep superoxide and hydroxide in the matrix side of the membrane interface, thereby enhancing phosphorylation yields.

• The new paradigm seeks only a simple functionality of Complex V (Manoj, 2017), which could be easily justified by its known structure. The new model does not seek Complex V to have TMP-tapping or harvesting and rotating abilities. A simpler operation of the γ stalk could permit proton inflow. The function of highly conserved and negatively charged β-DELSEED loop could be to recruit the protons coming in through $F_o$ subunit or to dispel a negatively charged part of the adduct in the "enzyme packaging of ATP". It is highly unlikely that the DELSEED "loop" (a structure with little self-stabilization energy) could serve as a rigid fulcrum for a purportedly rotating pivot of the γ stalk. A helix would have perhaps better served such an intended purpose. Further, the RCPE hypothesis has no role for 6 ADP binding sites on the $F_1$ subunit, whereas the murburn explanation is fully accommodative of this "rediscovery" of two ADP binding sites per β-subunit (Daniel and Zippora, 1984).

*Direct implications for DROS in physiological/phosphorylation biochemistry:*

\# If the new hypothesis were true, actively respiring and ATP synthesizing mitochondria could show an increase in DROS levels. This projection has been ratified in the physiological setups wherein DROS production was negligible when no active synthesis occurred but net ATP synthesis was directly correlated to increasing levels of DROS (Nicholls, 2004).

\# The new mechanism requires/predicts that DROS production should be directly correlated to the *in situ* concentration of oxygen and reduced substrates. This is in fact found to be a valid physiological premise (Murphy, 2009).

\# As per the new concept, each one of the respiratory complexes must be capable of generating DROS independently because the system is not a "sequential ETC". In fact, DROS are generated "enzymatically" by Complexes I through IV (Bleier and Dröse, 2013;

Dröse, 2013; Grivennikova and Vinogradov, 2006; Ksenzenko et al., 1992). This eventuality also agrees with the redox potential values of centers in Complexes I through IV.

# Non-specific phosphorylation of hydroxyl groups in amino acids with heme-flavin containing biological systems is direct support for DROS serving as couplers in phosphoryl ester synthesis (Blaurock et al., 1976; Spudich and Stoeckenius, 1980a; Spudich and Stoeckenius, 1980b). Further, involvement of phosphate ions in radical chemistry has been noted in mXM systems (Reinke et al., 1995) and an "ion-radical" mechanism for ATP synthesis has been a long-standing proposal (Buchachenko et al., 2010; Yasnikov et al., 1984). The observation of rare phosphorylation of hydroxyl groups of xenobiotics also testifies to DROS machinery's ability to carry out phosphate group transfers (Mitchell, 2016).

# The fact that multiple $^{18}O$ atoms (from water, the solvent) get rapidly incorporated into a single ATP phosphate (Cohn, 1953) is a strong support for radical mechanism. In photophosporylation, two $^{18}O$ atoms from Pi and one $^{18}O$ atom from water was found incorporated into ATP (Avron et al., 1965). Schemes proposed in (a1) through (b4) etc. explain the $^{18}O$ labeled experiments' outcomes. The fact that the phosphoryl group transfer chemistry was inhibited by dinitrophenol (Allison and Benitez, 1972; Cohn, 1953) in a parallel system and ATP synthesis is affected by plant phenols (Nakanishi-Matsui et al., 2016) is yet another strong evidence for the obligatory involvement of DROS in interfacial phosphyorylations.

# Involvement of the radical stochastic scheme dictates a variable and non-integral stoichiometry. The new reaction scheme predicts maverick effects upon varying redox active molecules' concentrations. Such effects of uncouplers and additives have been ratified in naturally phosphorylating systems (Avron and Shavit, 1965; Watling-Payne and Selwyn, 1974).

\# It should be possible to demonstrate ATP synthesis with a DROS generator and ATPase alone, in a closed vesicular system. The famed Racker's experiment and recent demonstration of reconstituting ATP synthesis activity in vesicles with only oxidase + Complex V are supportive of the new mechanistic concepts (Otrin et al., 2017).

• In chloroperoxidase catalyzed reactions, efficient chlorination is seen only with mM levels of chloride and this enzyme-sponsored reaction takes place "outside the active site, by diffusible species" (Manoj, 2006; Manoj and Hager, 2008). This mechanistic analogy is relevant because mitochondria houses millimolar levels of phosphate.

• Contrary to the prevailing perceptions (which see DROS as chaos infusing agents), recent research has established that enzymes need not get easily denatured by DROS and the former could use DROS to give specific substrate turnovers *in situ* (Manoj and Hager, 2001; Manoj et al., 2016c; Parashar et al., 2018). The fact that mitochondrial CYPs utilize the mitochondrial-membrane housed protein machineries for their activity *in situ* is another support for the application of oxygen-centric mechanism to mOxPhos. For diverse complexations, chlorinations and hydroxylations, a stable two-electron reaction product can be formed through one-electron routes (without wreaking havoc) in situ and such reactions are essentially involved in routine redox metabolism (Manoj, 2006; Manoj et al., 2010a; Manoj et al., 2010b; Manoj et al., 2016a; Manoj and Hager, 2008; Manoj et al., 2016b; Manoj et al., 2016c; Manoj et al., 2016d; Parashar et al., 2018).

\# If DROS is obligatorily required for physiological functioning, one would expect shorter life spans for animals with high metabolic rates (because with higher DROS, the side-reactions also are higher). This is in fact the statistical case in nature (Olshansky and Rattan, 2009).

\* The arguments professed herein can be further confirmed by employing- (a) simple chemical controls and reductionist approach, within controlled-water systems like

normal/reverse micelles, with DROS generating/stabilizing systems involving systems with Mg / Fe / peroxide / NADH / superoxide / pyrithione-photolysis (as a hydroxyl radical source) and substrate analogs.

* It is opportune to recollect that amphipathic DROS modulating molecules (like vitamin E and fattyacyl vitamin C) and trans-membrane helix containing redox-active enzymes (like horseradish peroxidase) inhibit membrane-embedded electron transport and redox metabolism in mXM system; whereas their soluble functional analogs (trolox, ascorbate, and superoxide dismutase) don't inhibit (Manoj et al., 2010a; Manoj et al., 2010b; Manoj et al., 2016a; Manoj et al., 2016b; Manoj et al., 2016c; Manoj et al., 2016d; Parashar et al., 2014a). A similar set of reactions could be repeated here with mitochondrial system or a minimalistic model like the Racker's/Otrin's experiment. In conjunction, the inhibitory capacity of retinol, retinal and retinoic acids and their esters could also be compared. [Jagendorf's experiment can be re-done by equilibrating mitochondria at pH 6 and then providing an external buffer of pH 4. The rates obtained above could be compared with an experiment where the mitochondria are equilibrated at pH 8 and then exposed to a buffer at pH 6. Such proton-gradients would be inadequate to power ATP synthesis (Manoj, 2017) per the chemiosmosis hypothesis even from the initialized state. But yet, we may observe a net equilibrium-driven ATP synthesis, indicating the inapplicability of chemiosmosis hypothesis.]

* ATP formation can be traced/quantified at $K_d$ levels of cyanide in freely respiring mitochondrial system. The new paradigm predicts a cessation of ATP formation in the physiological system but RCPE hypothesis warrants a stoichiometric lowering. In Racker's experiment, cyanide should not bring in any problem, as per the RCPE hypothesis. But the new hypothesis predicts inhibition of ATP synthesis.

* When a machine runs, it may make sounds or generate smoke. It is not the sound or smoke that runs the machine. It can be imagined that if a lot of sound/smoke is produced, the

machine may be working harder, but it is not the greater sound/smoke that makes the machine work harder. Therefore, with this analogy of sound/smoke to TMP (or even to the "detectability of DROS"), we could speculate on the working logic of mOxPhos. The mitochondria to work even without a pH gradient, at low proton concentrations and without high trans-membrane potentials. This projection can be ratified in simple experimental systems.

*Accounting for cyanide toxicity:*

• Per the RCPE explanation, cyanide inhibits Complex IV and therefore, is detrimental to ETC and proton pumping activity. KCN is lethal when consumed orally at ~1.3 mg/kg body weight. This means that for an average human (weighing ~70 kg), a concentration of 20 μM cyanide is lethal (assuming that average body has an equivalent density as that of water). *In situ*, the effective concentration of cyanide would be far lower than what is administered, owing to loss through gut retention or excretion. Cyanide, being an asymmetric species, would be relatively less hydrophobic than oxygen, and therefore, oxygen would most probably out-compete it within the phospholipid membrane. Further, the $K_d$ values proffered by cyanide binding at the heme center approaches ~$10^{-3}$ M for most hemeproteins under *in vitro* conditions, and it is highly unlikely that *in vivo* conditions would drastically change these values. It must be remembered that the anionic form binds to Fe-centres. The $pK_a$ of HCN is 9.4, at least two units higher than physiological pH. Why then, such a highly evolved oxygen-binding machinery of a plasma-membrane embedded Complex IV loses out even to trace quantities of cyanide (and other multi-atomic species like CO and azide)? Why does micromolar levels of an agent like cyanide mess with the respiratory logic? The RCPE hypothesis offers no explanation.

Since a thermodynamic pull cannot be exerted from Complex IV bound oxygen, one can gather that electron moves across favourable gradients via a very feeble thermodynamic push or is subjected to equilibrium within the Complex I/II – CoQ – Complex III system. Therefore, protons would be pumped out via Complex I & III even if Complex IV is blocked by cyanide (at least, in the first cycle). If the RCPE hypothesis were operative, an aerated and NADH replenished mitochondrial system should show some detectable ATP formation (in the presence of cyanide) because it is only the protons that sponsor Complex V activity. But the fact is that inclusion of cyanide fully inhibits ATP synthesis within mitochondria. This eventuality is seen when there could be several electron acceptors within the mitochondria (including the freely available oxygen, that could take electrons at say, Cyt $b_H$ of Complex III), enabling the "ETC" to recycle via Complex I and/or III (which could be analogous to the cyclic and pseudo-cyclic photophosphorylation cycles in thylakoid ETC).

Cyanide has low affinities for the oxidized Complex IV and for the reduced species, it has a comparable $K_d$ with respect to oxygen (Antonini et al., 1971). Therefore, toxicity of low concentrations of agents like cyanide was better explained in diverse heme-enzyme systems (heme-histidylate, heme-thiolate and heme-tyrolate) by a non-active site paradigm (Parashar et al., 2014b). Recent works have shown that under low concentrations and low mobility scenarios (akin to physiological conditions), species like azide and cyanide do not function as Fe-binding inhibitors but end up serving as "pseudo-substrates" (Andrew et al., 2011; Manoj et al., 2016a; Manoj et al., 2016b; Manoj et al., 2016c; Manoj et al., 2016d; Parashar et al., 2018; Parashar et al., 2014b). Therefore, cyanide kills at low doses not owing to binding at Complex IV, but because of messing the DROS (superoxide) equilibriums in milieu, per the following reaction.

$$CN^- + O_2^{*-} \rightarrow CN^* + O_2^{2-}$$
$$CN^- + OH^* \rightarrow CN^* + OH^-$$

$$CN^* + CN^* \rightarrow (CN)_2$$
$$(CN)_2 + 2OH^- \rightarrow CNO^- + CN^- + H_2O$$
-------------------------------------------------------------------------
**$CN^- + OH^* + O_2^{*-} + OH^- \rightarrow CNO^- + O_2^{2-} + H_2O$**

Thus, we can see that cyanide can effectively form two-electron sinks in the overall one-electron scheme. For the first time, we thus have a tangible explanation for the toxicity of low concentrations of cyanide.

* The chemical relevance of DROS can be traced by the incorporation of radio-labeled carbon in cyanate, after exposing an experimental system to sub-lethal amounts of radio-labeled cyanide. Else, $^{18}O_2$'s labeled atom incorporation into cyanate could be traced, when cyanide is presented to mitochondria.

*Reasoning other inhibitions and uncoupling:*

• The known mitochondrial toxin of vitamin A (de Oliveira, 2015) is not reported to have any proton-pump inhibiting or pmf uncoupling activities but it can surely be deemed a DROS modulator at the phospholipid interface. Therefore, the new mechanism explains the toxicity of Vitamin A to mitochondrial metabolism.

• Di-substituted phenolics serve as uncouplers (read- inhibitors of reaction of interest) in both mXM and mOxPhos systems (Manoj, 2017; Parashar et al., 2014a; Parashar et al., 2018). As inhibition remains a key to understand mechanistic implications, the essential role of DROS in mOxPhos metabolism is underlined.

• Complex I and II inhibitors like rotenone and stigmatellin inhibit the electron "drip-line flow" by clogging the immediate electron sink of the protein complex in the trans-membrane region. Without this, there is no "static discharge" of electrons from the flavin to the distal FeS center, as a result of which electrons cannot be ferried away from the protein complex.

• The new paradigm explains the toxicity of cationic lipids used for drug delivery (Aramaki et al., 2001). The positively charged lipids would mess with the negatively charged radicals in membrane/milieu.

• Heat generation by chemical uncouplers (interfacial DROS modulators) and uncoupling protein (which has positively charged amino acid residues within their trans-membrane helices (Klingenberg and Huang, 1999) that could modulate DROS) is yet another support to the new paradigm.

• The new understanding rationalizes why certain uncouplers (like dinitrophenol) or ionophores (like valinomycin) permit electron transfers ($O_2$ consumption or depletion of NADH) but not ATP synthesis, with intact or disrupted (by shear forces or by detergents) mitochondrial system. Further, ATP synthesis does not occur (with or without uncouplers) in broken mitochondria and this could be because effective murzone formation (possible only in a low-proton, low-water activity regime) is prevented. As a result, the radicals formed do not get to attack phosphate, but can effectively react with each other (because lot more protons are present in the free solution) in a futile manner. This process serves as a secondary two-electron sink (pulling drive) and NADH can thus get effectively depleted and oxygen can get consumed. However, in the proton-limited intact mitochondrion, ADP + Pi is needed to serve as the electron "diverting agent" (leading to $O_2$ consumption or NADH depletion, and formation of water) because under such a limiting regime, the reaction can occur to proceed to completion only with the protons available from these substrates and their linking thereafter to yield a stable product. This idea of "redox pull" has been demonstrated in the mXM system. The new paradigm could also explain why inclusion of uncouplers alone (without ADP + Pi) cannot consume oxygen / deplete NADH in intact mitochondrial membrane system, whereas the RCPE hypothesis falls absolutely flat. If "ETC" occurs in mitochondrial fragmented system, why should it not occur in the intact system is a question that RCPE cannot answer.

• Now, we can understand why Complex V's $F_o$ subunit binders like venturicidin and oligomycin (and trans-membrane porting system blockers like bongkrekate and atractyloside) inhibit ATP synthesis in the steady-state. If facilitated chemostasis (pH maintenance) does not occur, ATP formation in the closed system is limited by reactivity and equilibrium considerations within. ATP and water must be formed and taken out for the reaction to proceed. (Analogy- if the chimney gets clogged, the firewood does not burn in the hearth.)

*Solving other long-standing conundrums:*

\# A non-integral and variable stoichiometry is naturally expected in a "stochastic operating system" such as the murburn concept (Manoj et al., 2016a; Manoj et al., 2016b; Manoj et al., 2016c; Manoj et al., 2016d; Parashar et al., 2018; Venkatachalam et al., 2016). This has already been experimentally ratified (Brand and Lehninger, 1977; Hinkle, 2005).

\# The new paradigm projects an intrinsic loss of redox equivalents even in the absence of "intended substrates". The "metabolic leaks" and oxygen uptake rate differences ("metabolic states of mitochondrial research") could be potentially explained by the "non-specific reactivity and redox/thermodynamic pull operation" of the new understanding (Manoj, 2017). The analysis of ETC shows that the prevailing model can at best serve as a back-up to a more spontaneous and faster scheme of electron transfer reactions. Therefore, the leakage of electrons before addition of ADP+Pi (in the "metabolic states" of mitochondrial research) is primarily via the combination of the new and old ET routes. The extra consumption of oxygen upon the addition of ADP+Pi results purely owing to the murburn-dripline relay.

• The new perspective better explains the "stress" at both low and higher concentrations of oxygen and the proton-deficiency/dynamics is also now explicable in the mitochondrial system.

• The new paradigm explains how simple unilamellar vesicles formed from inner mitochondrial membranes could mediate electron transfers coupled to ATP synthesis. RCPE paradigm would not work in such a reconstituted system as it would lack the obligatory Cyt. *c* mediated inter-membrane electron delivery system, an obligatory requisite for Complex IV to form water.

• The amount of ATP formed would be lower with succinate as the starting material because of the poor efficiency of superoxide formation starting from a relatively higher redox potential (and also by the fact that this system is not proton deficient).

• The disparity in $K_d$ & $K_M$ (in this unusual case, $K_d >> K_M$ (Orii, 1988)) for Complex IV's interaction with oxygen can be reasoned with the new paradigm (Venkatachalam et al., 2016).

• The anoxic electron transfer rate between two "heme *a*" of Complex IV was found to average at $10^{-1}$ s$^{-1}$, whereas only the oxygen-pulsed system showed rates of $10^3$ s$^{-1}$, approaching the physiological electron transfer values (Orii, 1988). This observation cannot be accounted by Marcus' outer sphere electron transfers. If we see DROS as the electronic conduit (as murburn concept advocates), this process can be easily explained.

• The new mechanism affords a strong chemical coupling logic between electron transfer and phosphorylation, explaining the enhanced utilization of oxygen in the presence of ADP+Pi.

• Since the reaction is a radical chain type of process, it explains the low "activation energy" barrier for ADP phosphorylation in physiological milieu.

***Evolutionary and holistic perspectives:***

\# The operating logic of the oxygen-centric concept invokes a hearth or nuclear reactor type of working principle (Manoj, 2017). While the hearth type logic is commonly seen, the spontaneous formation of a nuclear reactor would be rare. The documentation of a naturally formed nuclear reactor on earth (Meshik et al., 2004) is testimony to the feasibility of spontaneous evolution and operative sustainability of the "stochastic" or "nuclear reactor logic" explaining mOxPhos.

\# When life evolved in the primordial times, highly specific affinity-based reactions based on topological and electrostatic complementation could not have existed. The fact that electrons can be put in and taken out through several man-made molecular/radical species (Hauska, 1977) vouches for a non-specific one-electron reaction paradigm.

\# The new proposal raises the yield barrier of RCPE (ATPs per NADH/succinate), thereby ratifying the evolutionary selection of "oxidative phosphorylation" as the preferred mode of ATP synthesis. This prediction has already been verified by several researchers (Hinkle, 2005).

• The chemical nature and availability of the key reactants in milieu agree with the new mechanistic scheme. Oxygen (available at $> 10^{-4}$ M) could move $10^2$ nm/µs (Milo and Phillips, 2015) via simple diffusion, affording viable reactivity and electron transfers. Protons at $5 \times 10^{-8}$ M can be available at a trans-membrane rate of ~$10^3$ s$^{-1}$, which is the overall experimentally observed (limiting) rate of mOxPhos. Thus, the new explanation fits the overall ET bill of mOxPhos.

• Cellular metabolic processes evolved for a controlled one-electron oxidation of NADH (a stable biological small molecule with a low redox potential within the physiological window of -400 to +800 mV) by oxygen, the latter going through superoxide-peroxide-hydroxyl radical/hydroxide ion, finally forming water. NADH became the choice of terminal reductant owing to its proton-limiting structure that could incorporate the two electrons (and its ability to give one-electron equivalents also (Zielonka et al., 2003)). Succinate could be a "governable" connecting link to the overall stochastic operative logic of mOxPhos.

• The new paradigm explains why leaching of Cyt. *c* leads to cellular apoptosis (Ow et al., 2008) in multi-cellular systems. Multi-cellular organisms could have "learned" that once the mitochondrial membranes were compromised, it was in their interest to "lose" the cell because the flavins and metal centers would otherwise wreak havoc.

\* An evolutionary-lineage analysis of respiratory complexes may give evidence of DROS modulation machinery's progression and specific adaptation through the time-line of evolution of species.

*Conclusions*

By the explorative investigations and critical inquiry carried out herein, it is inferred that proton-aided ATP synthesis by the RCPE mechanism is inadequate to account for the steady-state mitochondrial physiology. The study points out binding-site and channel based evidences, directly implying DROS mediated catalysis in phosphorylations. The inhibitions/uncouplings offered by interfacial DROS-modulating agents is a strong support for the importance of diffusible radicals. The new oxygen-centric proposal explains quick death in the absence of oxygen and the lethality of low amounts of cyanide. Further, we can avail direct chemical reaction connectivity between NADH oxidation and phosphorylation. Therefore, oxygen is touted to be the essential electronic conduit and catalytic agent in cellular respiration, as it is via DROS that coupling (ATP+$H_2O$) and uncoupling (Heat+$H_2O$) occurs. In the context of metabolism, several global cultures classify food into "hot" and "cold" categories (Bonder et al., 2002). Considering that these perceptions have statistical basis, one could theorize that the mOxPhos and mXM metabolic routines could get significantly routed towards uncoupling/heat generation. Therefore, a new perspective of "pharmacokinetics-dependent pharmacodynamics" is opened up. While Linus Pauling's fascination for vitamin C (for its antioxidant roles) is documented, James Watson had recently proposed that DROS could be useful for preventing several pathological states (Watson, 2014). Therefore, it would be justifiable to determine the details of criteria for DROS to play Dr. Jekyll and Mr. Hyde. Presenting several arguments, this work is the third distinct case where DROS have been implicated to effect sustainable physiological (redox metabolic) outcomes. Though the new explanation for mOxPhos places several elements in a more coherent perspective, some roles of the respiratory complexes and the overall details of phosphorylation chemistry remain uncharted. Therefore, the perspectives offered herein could give rise to new avenues for scientists to explore. It is speculated that the new mechanistic explanations could be relevant in a wide variety of physiological processes, including photosynthesis, cellular homeostasis and a plethora of physico-chemical signal-transduction schemes.

*Acknowledgments & Disclaimers:* This work was powered by Satyamjayatu: The Science and Ethics Foundation. KMM dedicates this manuscript to the fond memories of Late Lowell P. Hager (Member, NAS-USA and University of Illinois at Urbana-Champaign, USA). Vivian David Jacob and Adhiatma Therat are acknowledged for assistance with manuscript-proofing and artwork, respectively. Our forthright critique of highly recognized ideas does not connote any disrespect towards pioneers/peers who meticulously work(ed) in the field. We only differ with the interpretations of the highly trustworthy data they generated. The authors have no conflicts of interests to declare with respect to the context.

**Author contributions:** KMM conceived the ideas, planned the experiments, reasoned the findings and wrote the paper. AP carried out the dockings, cavity analyses, reported data and collated the bibligoraphy using Endnote.

# Supplementary Information

# Unveiling ADP-binding sites and channels in respiratory complexes: Validation of Murburn concept as a holistic explanation for oxidative phosphorylation

*Kelath Murali Manoj\* & Abhinav Parashar*

\* Satyamjayatu: The Science and Ethics Foundation,
Kulappully, Shoranur-2, Palakkad District, Kerala State – 679122, India.

**Contents:**

**Item 1:** Procedural and visual details of in silico docking of various proteins and exploring the channels within the respiratory complexes

**Item 2:** A critical view on Boyer's rotary ATP synthesis proposal

**Item 3:** Analysis of the prevailing ETC model

**Item 4:** A comparison of the working logics of RCPE and murburn concept perspective of mOxPhos

*References*

# ITEM 1

Preliminary screening with a spacing of 0.85 Å using Autodock's Lamarckian genetic algorithm (Table A1 below) accurately detected the unique binding site of biotin on streptavidin (Figure A1). However, the $K_d$ values were markedly lower compared to the ones reported in literature. But the unique ADP-binding sites of known positive controls like phosphoenolpyruvate dephosphorylase (pyruvate kinase, $K_d \approx 10^{-5}$-$10^{-4}$ M; Figure A2) and phosphocreatine dephosphorylase (creatine kinase, $K_d \approx 10^{-6}$-$10^{-5}$ M; Figure A3) agreed with values quoted in literature. Further, non-specific interaction controls (chloroperoxidase-ADP and amylase-ADP) gave $K_d > 10^{-3}$ M ranges. Therefore, this "validated crude docking" protocol was used as a "screening filter" for detecting putative ADP binding sites on the various (super)complexes (Table A2). Once the sites were identified, re-docking with well-defined grids was carried out at a spacing of 0.375 Å, and these data are reported in the main manuscript. (However, it is opportune to disclaim that the *in silico* $K_d$ cannot be taken as an "authoritative index of binding affinity. Herein, a $K_d < 10^{-4}$ M is merely used as an identification tool for a putative/generic ADP-binding site.) Figure A4 shows another view of Complex I's extended channel's connectivity to the solvent. Different views of Complex III and its channels/interaction with ADP are presented through Figures A5 - A8 and the appended movie file (Video A1).

*Table A1: Controls employed in the preliminary docking studies with 0.85 Å spacing*

| Protein [pdb]-ligand | Source | Type | Cluster(s), Affinity | $K_d$ |
|---|---|---|---|---|
| *Streptavidin (1MK5)-Biotin* | *Streptomyces avidinii* (Hyre et al., 2006) | Protocol control | 1, high | 10 μM |
| *Pyruvate Kinase (3D2R)-ADP* | *Homo sapiens* (Wynn et al., 2008) | Positive control | 1, modest | 39 μM |
| *Creatine Kinase (1G0W)-ADP* | *Bos taurus* (Tisi et al., 2001) | Positive control | 1, high | 2 μM |
| *Amylase (1SMD)-ADP* | *Homo sapiens* (Ramasubbu et al., 1996) | Negative control | 1, low | 2.2 mM |
| *Chloroperoxidase (1CPO)-ADP* | *Leptoxyphium fumago* (Sundaramoorthy et al., 1995) | Negative control | 1, low | 7 mM |

## Streptavidin (1MK5)-Biotin

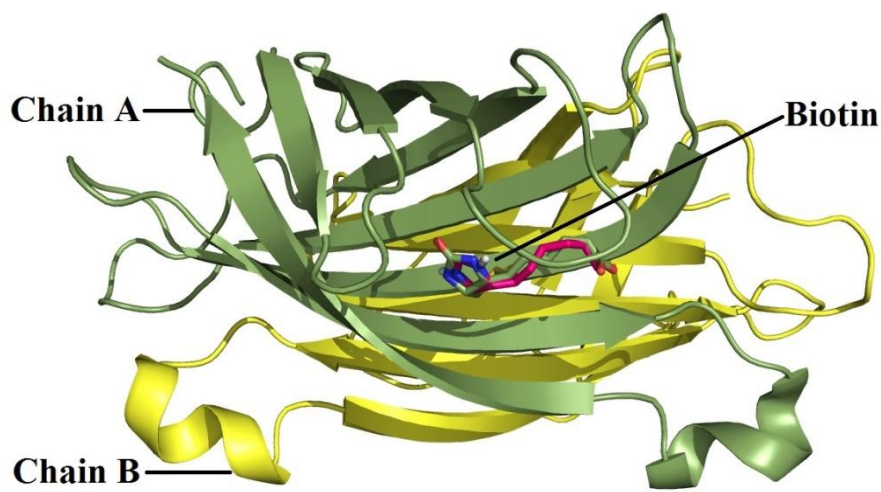

**Figure A1: Protocol control.** *The crystal structure binding of biotin is overlaid with dock biotin conformer of lowest binding energy (magenta). Streptavidin is shown in its physiological dimer state with the two monomers colored in olive green and yellow respectively.*

## Pyruvate Kinase (3D2R)-ADP (Positive Control)

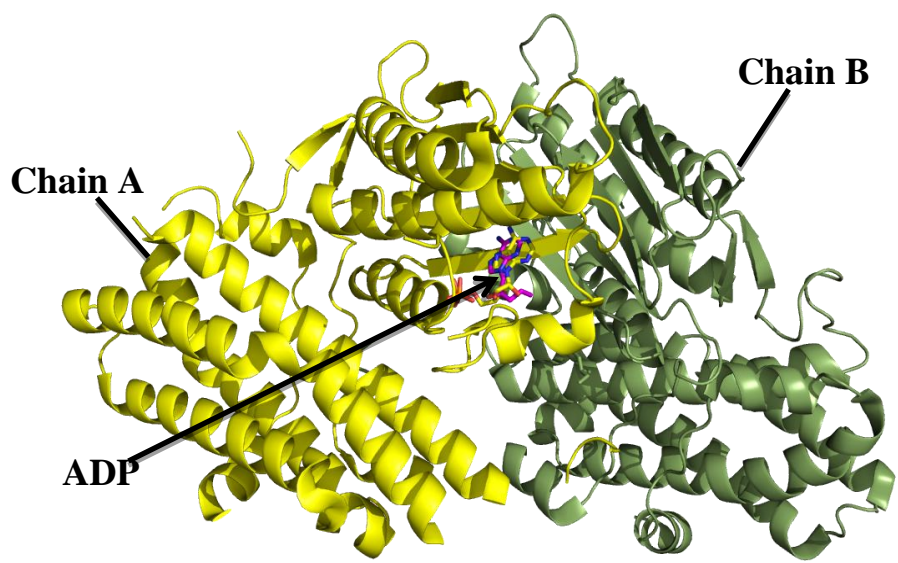

**Figure A2: Positive control.** *Superposition of docked ADP with the crystal structure demonstrates the same location of binding, as in silico. The two monomers of this dimeric protein are depicted in olive green and yellow.*

## Creatine Kinase (1G0W)-ADP Interactions

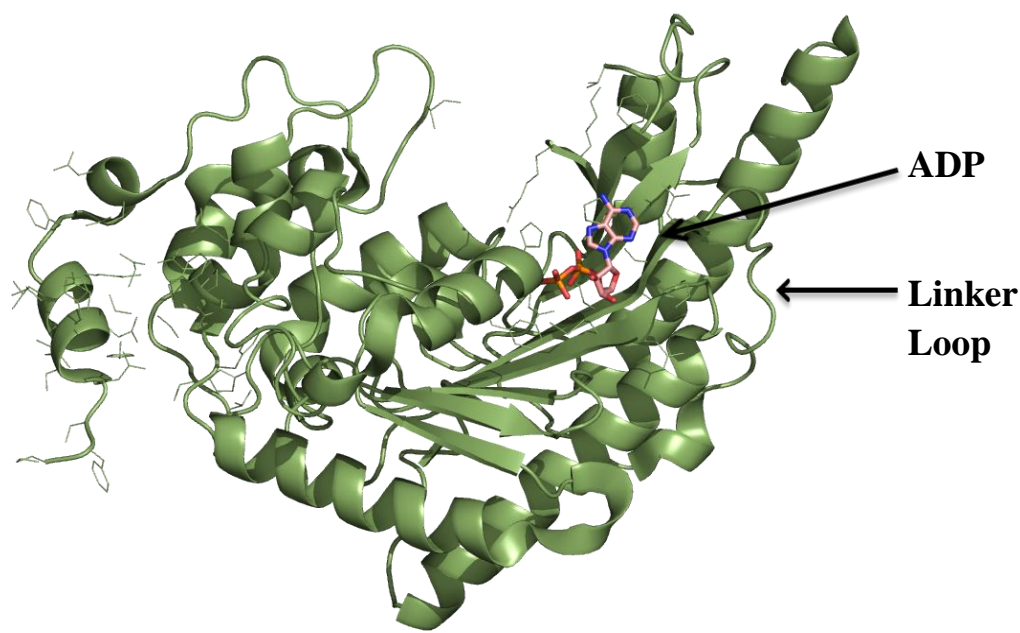

*Figure A3: Positive control. The protein shows high affinity towards ADP. The binding site, as reported in crystal structure (adjacent to the linker loop,) is correctly predicted by the docking analysis.*

*Table A2: Details of proteins used for docking/cavity analyses*

| *(Super) Complex* | *pdb* | *Source* | *Reference* |
|---|---|---|---|
| **Complex I** | 5LDW | *Bos taurus* | (Zhu et al., 2016) |
| **Complex II** | 3SFD | *Sus scrofa* | (Zhou et al., 2011) |
| **Complex III** | 2A06 | *Bos taurus* | (Huang et al., 2005) |
| **Complex IV** | 1OCC | *Bos taurus* | (Tsukihara et al., 1996) |
| **Complex V** | 1H8E | *Bos taurus* | (Menz et al., 2001) |
| **Respirasome** | 5GPN | *Bos taurus, Sus scrofa* | (Gu et al., 2016) |

# Complex I: Solvent channels, cavities and ADP interaction

Side view of the matrix region of Complex I:

Solvent channels in matrix region of Complex-I leading to Fe-S and Flavin centers indicate probability for access/formation of oxygen/DROS.

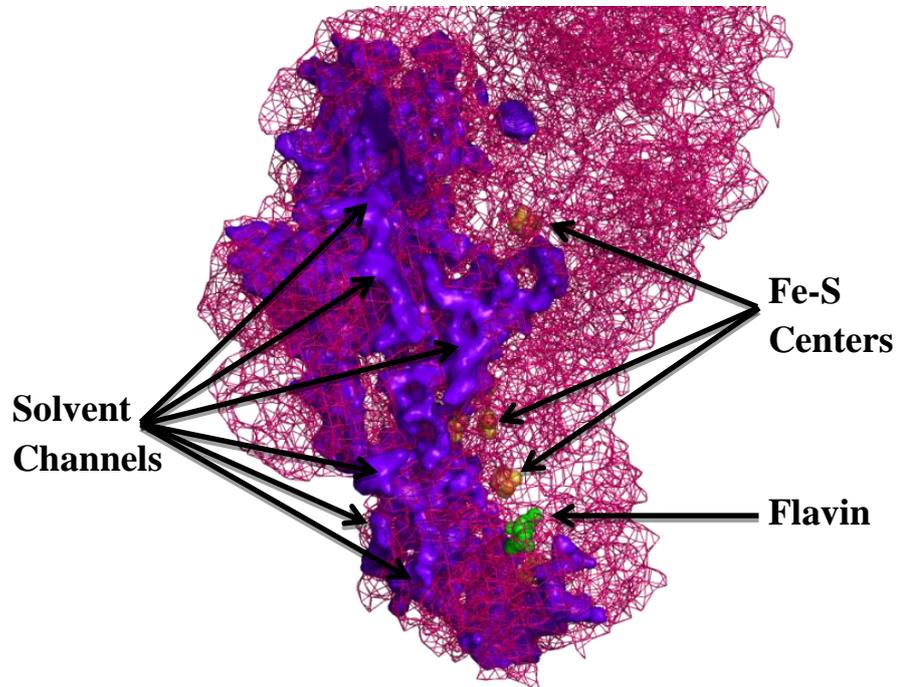

**Figure A4: Solvent channels of Complex I**

## Complex III: Solvent channels, cavities and ADP interaction

Surface view of whole Complex III homodimer illustrating the channels and cavities along with the position of Heme $b_H$ and ADP binding sites. Since the complex is a dimer, a total of four ADP binding sites (two each on one side of the dimer) are available. Only one site is visible in this orientation and rendering. Another site is hidden under the shadow.

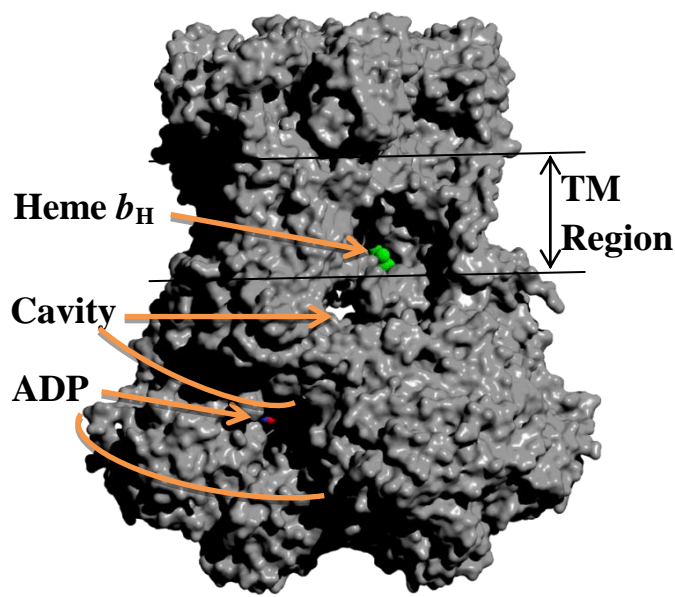

**Figure A5: Complex III, sideview, with transmembrane region**

Tilted side view of the above represented image: Here we can see the two ADP binding sites on one face of the complex.

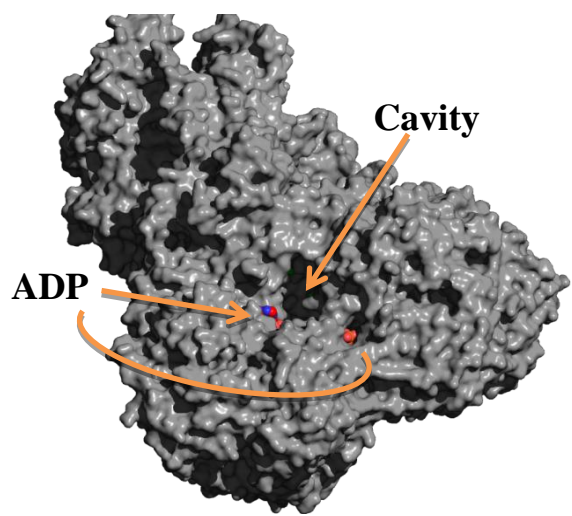

**Figure A6: Tilted iew for the surface of Complex III**

**Internal channel** that connects the big cavity of Complex III homodimer in the matrix region and **Heme $b_H$** in the transmembrane region.

Hydrophilic surfaces are shown in blue and hydrophobic in red.

Image created using CAVER 3.0.

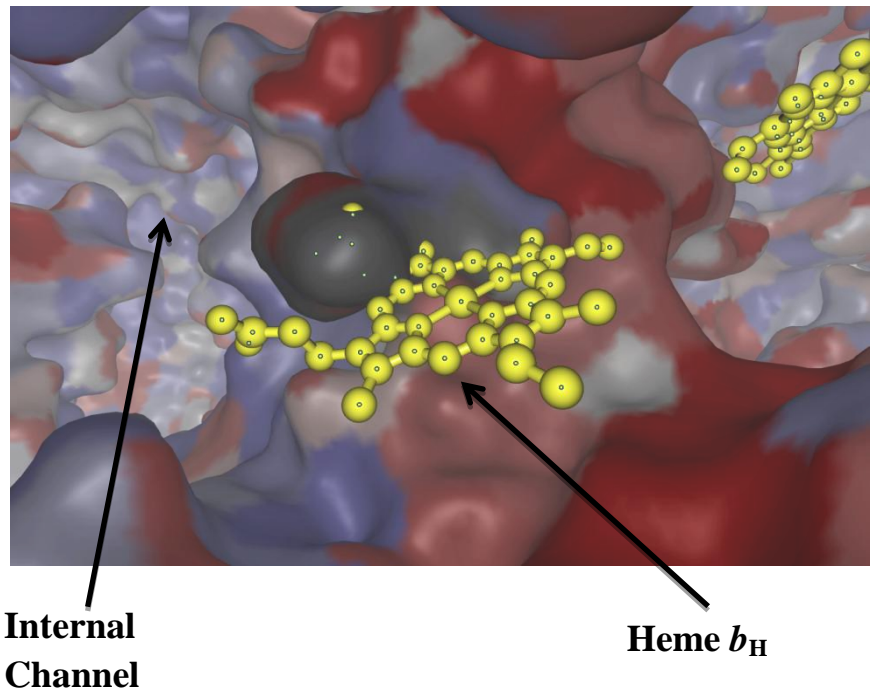

**Internal Channel**

**Heme $b_H$**

**Figure A7: Internal channel in Complex III near heme $b_H$**

A couple of views of Complex III homodimer represented with the different conformers of ADP interaction as predicted by AutoDock 4.2

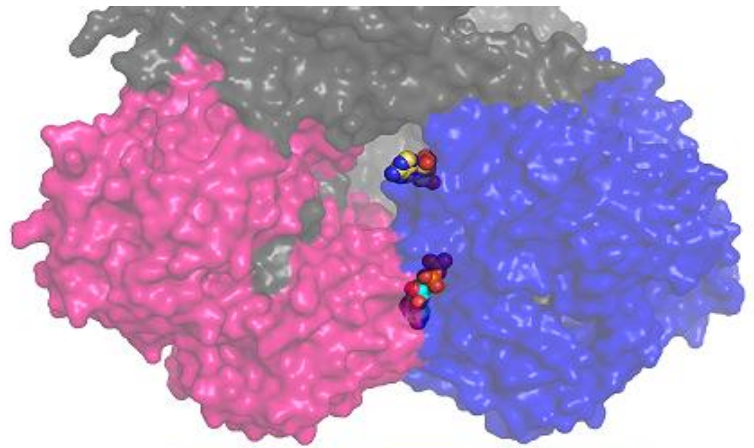

**Bottom view**

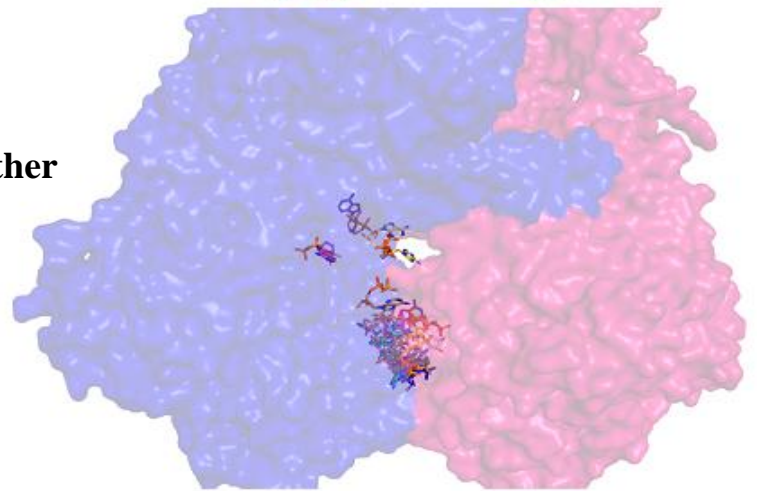

**Another view**

The position of ADP adjacent to a solvent cavity/channel is evident.

**Figure A8: ADP binding clusters on Complex III**

# ITEM 2

In free solution, the rate of hydrolysis exceeds the rate of ATP synthesis by a factor of $\sim 10^5$. This is reflected in the fact that in physiological milieu, thermodynamics dictates a significant free energy change of about -30 to -35 kJ/mol for ATP-hydrolysis. (That is- hydrolysis is highly favoured over esterification.) Therefore, it is evident that energy must be expended to synthesize ATP.

Boyer's first postulate is that the motor of Complex V is driven by the real-time power generated from TMP. [This was presumed to come from proton disparity/movement, which in turn, was supposedly driven by the energy derived by the oxidation of ETC fuel molecules, NADH and succinate. Such a supposition has already been discredited (Manoj, 2017).] Energy is need for moving the $F_o$'s *c*-drum/γ-shaft, and thereby leading to the detachment of the formed/bound ATP at the $F_1$ module. This is equivalent to making the enzyme lose its affinity for ATP. But then, how is the ATP formed there, in the first place? The second postulate that caters to this requisite is- yet another high affinity site on Complex V must simultaneously bind (ADP + Pi) with a much higher efficiency than it could bind ATP. That is, in the cyclic synthesis process, the same binding sites go through a high affinity for ADP versus high affinity for ATP cycle. If Complex V functions reversibly within the system, the affinities for ADP and ATP should be quite comparable. But the *in vitro / in situ* works suggest that the esterification routine overcomes the rate of hydrolysis reaction and Complex V has $>10^7$ folds higher affinity for ATP (than ADP). Boyer's postulates are not supported by simple evidence or straight-forward scientific deductions. It is our opinion that such proposals gathered traction only because Mitchell's chemiosmosis proposal was around. Let's recollect that Boyer had erred when he proposed "Phosphohistidine" as the enzyme-linked high energy phosphorylating intermediate (Boyer, 1963) when Slater's hypothesis was in vogue (Slater, 1953) and Mitchell's idea had not caught on. Nobel laureates, like all human beings, are not infallible (Allchin, 2008).

A simple way to question Boyer's mechanism is by looking at the way the γ stalk/shaft functions in a rotating mode. The movement of the γ shaft cannot work as a "pushing out agent of ATP" in one direction and "pushing out agent of ADP + Pi" in the other. That is- changing the direction of the shaft does not change affinities. Stated otherwise, Complex V cannot work as a hydrolyzer in one direction and esterifier in another. The non-provable "conformation changes" and "non-harnessable" TMP cannot be quoted to serve as "marvellous tour-de-force" for overcoming the mechanistic/thermodynamic odds that RAS hypothesis faces in physiology. Regardless, let's understand the "rotary hydrolysis activity" first. In the hydrolytic function, the stalk may rotate because of the favourable ATP binding (and hydrolysis reaction that releases energy). In this mode, an ATP molecule present in the matrix at high concentrations can efficiently bind $F_1$, get hydrolyzed and this energy can lead to a conformation change and twirl the shaft. In the meanwhile, there is a good probability that the next ATP (because it has higher affinity and is found at a higher concentration than ADP) would be bound on the adjacent site and the hydrolytic process could go on if all the other structural mandates are met. At the inter-membrane side, the $F_1$ activity could relay a change in the $F_o$ domain, which could induce an inward movement of protons. This function would not need a rotary function, and just a "sliding/opening" kind of "conformation change" would allow protons to spontaneously/actively enter to neutralize the negative charges developed within.

But in the esterification mode, the causative is the binding of protons on the outside. The dynamics at the $F_1$ subunit must ensure that both ADP and Pi bind effectively at a precise instant the proton(s) trigger the $F_o$ module in the disconnected inter-membrane space. This seems a less probable proposition. In the best case scenario favoring the RCPE hypothesis, a surplus of protons entering from outside through $F_o$ module would keep rotating the γ shaft. Since the $F_1$ portion has high affinity for the single substrate ATP (and since ADP in matrix is significantly at lower levels than ATP), it will always out-compete the binding process of the two substrates (ADP and Pi) at the internal F1 binding site. Therefore, the ATPsynthase could only liberate pre-synthesized ATP as the γ stalk ploughs through the $F_1$ module.

It must be said that single-molecule experiments have little ways of ascertaining spatio-temporal relevance or kinetic/energetic viability of Boyer's ideas in realistic physiological setups. Such an experiment does not demonstrate that Complex V is "rotated" by an inward movement of protons either! *In situ*, there would be several requisites- The *c* subunits' cylinder (attached to the shaft) would have to hold on to the membrane and also rotate with respect to the membrane, imparting a torque that enables the γ shaft to move through the α-β dimer-trimers. Such movements in both directions must not destabilize the protein complex, and all α-β dimers should be equivalent, etc. (Structural elements do support these assumptions.) Considering that the inner mitochondrial membrane has bare minimal lipid content (only 20 %), it looks unlikely that the $F_o$ domain could find ample "fluidic-tethering" within such a "rigid membrane" (or hinge through another protein unit) and rotate at the same time. If these aspects are not comparable with the experimental setup (His-tag tethering of the $F_1$ module to a slide and noting the rotation of a bound actin filament to the γ shaft/ *c* drum), such an experiment cannot serve as an evidence to show that Complex V is a cyclic enzyme or a rotary synthase. Quite simply, there was little way such an experiment could have shown that Complex V is not a rotary enzyme. It is opportune to cite an analogy from the mXM system here. Elaborate studies published in reputed journals had "demonstrated" several modalities/structures within an "improbable" purported catalytic cycle of cytochrome P450 (Schlichting et al., 2000) or complex schema of CPR-CYP/Cyt. $b_5$ associations for inter-protein electron transfer (Jamakhandi et al., 2007). Such "unrealistic evidences" were "inapplicable" *in situ*, as the enzymatic system was demonstrated to recycle by more facile ways (Manoj et al., 2016). Another simpler analogy could be- in a circus ring, a lion would be made to jump through flaming loops, out of fear for its sustenance. In the wild, the lion makes its living in quite a different way.

Definitely, Complex V could potentially synthesize ATP inefficiently by an equilibrium assisted process when ATP is depleted in mitochondria (and that too, at very high levels of ADP and Pi; if we accept that Complex V's $K_d$ value of ATP is $10^7$ times lower than ADP) and when a significant external pH gradient is given. Such a process has been demonstrated [the famed Jagendorf experiment, (Jagendorf and Uribe, 1966)] and this equilibrium driven synthetic reaction would be different from the physiological steady-state ATP synthesis which occurs without a pH gradient and with excess of ATP. This is analogous to a typical hydrolase like lipase, which carries out the faster hydrolysis in normal micelles, but catalyzes the slower esterification in reversed micelles at low water contents (Han et al., 1987). The relevance of such type of a Complex V mediated reverse reaction in mitochondrial ATP synthesis can be safely side-lined, as it would be equivalent to having a juicer give the fruit, starting from pulp and juice. A juicer can give only juice and pulp from fruit; the fruit gets synthesized in nature by an alternate route. Using the juicer (in an altered configuration) to make the fruit from juice and pulp can at best give us a concentrate, but not the fruit. There is little concrete chemical or physical logic as to how the very same enzyme can reverse its affinities selectively, when the overall phenomenon involves at least three distinct phases (matrix, membrane and IMS) and the process involves the recruitment of several molecules/ions, etc. While some simpler aqueous phase reactions are easily reversible and can be manipulated by varying the concentration terms of reactants (thereby harnessing the equilibrium forces, as mentioned earlier for examples like lipase), Complex V's proposed working complexity does not permit reversibility.

Let's consider a truly reversible phosphorylating enzyme like creatine kinase. The data from a crisp review by McLeish and Kenyon serves well to demarcate the concepts involved, as excerpted in the Table 1 within the main manuscript. From this example, it is clearly evident that higher rates (first or second order) are obtained in the synthesis reaction for all isozymes A through D only because the $K_d$ for ADP is smaller than ATP for the enzyme. Therefore, in physiological scenarios, Complex V cannot mediate ATP synthesis if it has a higher affinity for ATP and even if the enzyme cycles through an equal number of high affinity for ADP + high affinity for ATP cycles.

Therefore, rotary ATP synthesis activity cannot prevail over hydrolysis with the known features of Complex V, under physiological scenarios. Even if we overlook the facts that- (i) mitochondria lack adequate protons to serve any outward proton-pumps, (ii) chemiosmosis principle was demonstrated to be a non-workable idea, (iii) physiology mandates a more efficient energy-harnessing process, etc.- and grant the rotary ATP synthesis proposal all consideration, Complex V or mitochondrial structure has no sophistication to afford physiological ATP synthesis via a rotary modality. Complex V's structure does not have any electro-magnetic induction principles or ferromagnetic components to tap into a supposed electric field that could be purportedly set up in the dynamic steady state. Also, if a continuous trans-membrane potential (with a defined polarity of inside negative and outside positive) is what drives the Complex V, how would it afford any temporal window for the any protein to go through a "native-excited-reorganize-native alternating cycle" scheme? Clearly, since the system is isothermal, the few number of protons' kinetic energy also cannot be tapped. Even if we grant a high level of sophisticated regulatory mechanism for mitochondrial functioning, we cannot derive "holistic/perpetual workability". The scenario is analogous to having two parties tug at a rope continuously in a small room. If the rope gets displaced, useful work is done but this cannot continue in steady-state because of dimensional limitations. On the other hand, if both parties pull with equal forces (but in opposite directions), then the system is in steady state but no work is done. Very importantly, one wonders how such a "miraculous" enzyme could ever have evolved and how it could spontaneously assemble on a routine basis. If the *c*-drum could spontaneously rotate with the stimulation of a proton, why should it bind to the "*a*" subunit?

Therefore, we can safely infer that chemiosmosis (movement of protons across the same membrane in a closed system) and rotary ATP synthesis (Complex V changing its affinity periodically for ADP and ATP; yet affording selectivity for ATP production) proposals are "non-workable machine" logics. Both chemiosmosis and rotary ATP synthesis demand highly deterministic sophistication from a non-modular, non-synchronized/staggered and non-sophistically regulated molecular system such as the mitochondrion. Further, the chemiosmosis-based ATP synthesis is inadequate to explain the energetics of cellular metabolism, as deduced from theory and as verified experimentally by several researchers (Manoj, 2017).

Let's now deal with the single molecule experiment that is often taken as a support for the rotary ATP synthesis proposal of Complex V. If we equate the proton-pump generated potential temporally versus the power needed to turn the ATP(synth)ase, we could make a quantitative analysis for testing the physiological relevance of rotary ATP synthesis. [To minimize complications, let's start with a mitochondrion possessing a single ATP(synth)ase.] Defining electric power as a product of voltage (0.2 V; sought by Mitchell and experimentally ratified by literature) and current (1.6 x $10^{-13}$ A; derived by assuming that about ~10 protons are present in mitochondria and these many protons give a flux of about at $10^4$ per second, [$10^1$ $H^+$ x $10^4$ $s^{-1}$] / 6.24 x $10^{18}$), we have the power (= V x I) equivalent to 3.2 x $10^{-14}$ $Js^{-1}$. (Let's call this the left hand term.)

From the literature on single molecule experiments (Tanigawara et al., 2012), the stepping torque (for the first 120° movement) for ATPase was determined to be ~38 pN.nm. Disregarding inertia, we can project that this torque must be generated 3 x $10^3$ $s^{-1}$ (because 3 x 120° = 360°; no. of rotations = $10^4$ protons moving in per second / ~10 protons per rotation). Since power = force x distance / time, the value of 1.14 x $10^{-16}$ $Js^{-1}$ is the power consumption for ATP(synth)ase. (Let's call this value the right hand term.)

Now, in steady state, the power of the TMP roughly equates with the power for the γ shaft's movement + power for the change of conformation of the ATP(synthase) bulb.

(Theoretically, both lead to the same outcome.) Then, calculation shows that if RCPE worked with 10 protons, only ~0.3 % (of the power that was generated) is spent for moving the ATPase stalk. If we assume that the number of protons solicited by RCPE is somehow present in the mitochondrion (amounting to a conservative number of >100,000), then only <0.00003 % is spent for moving the ATPase stalk(s). Now, where goes the rest of the energy involved/generated? [Now, if one tries to factor in the copious amounts of ATP present to the right hand side, the left hand term would then fall significantly short of the right hand term!] On the other hand, if we try to lower the current component by altering the proton flux to a lower number (say- 10 protons move out/in at $10^3$/s or $10^2$/s), it would give still give only 3 or 30 % of consumption of power generated. If we lowered the proton level to any lesser order, we would get to the scenario wherein the trans-membrane potential would fail to "power" the ATPase motor. (That is, the left hand term becomes smaller than the right hand term.) Therefore, we must seek a particularly narrow "aesthetic" proton concentration range and/or flux and/or protein density to justify the current component sponsored by protons. But if we do that, we cannot simultaneously meet up the demands of the experimentally observed pmf and protein densities (with the same proton/flux numbers, even if we accept a highly optimized RCPE model as discussed earlier (Venkatachalam et al., 2016)).

The predicament we are faced with for justifying RCPE is- only a portion of the energy derived from NADH oxidation can be used to pump protons out. There is no way that the physiological system can recycle the spent energy. Even if we (violate the fundamental laws of physics to) accept this proposal that proton re-entry is harness-able, only a portion of the "recycled energy" can be utilized. Even if we do that, the efficiency calculations of such protons' return don't afford any quantitative justifications for the comprehensive/dynamic physiological functioning. This scenario could be analogous to the following "one buck" accounting problem (which can be made out in many formats, of which I present one popular variant). The following entries are made in a guy's expenses' diary.

Bucks at hand = 50

|  | Expenses | Remainder |
|---|---|---|
| Drinks - | 20 | 30 |
| Snacks - | 15 | 15 |
| Stationery - | 9 | 6 |
| Transport - | 6 | 0 |
| Total → | 50 ≠ | 51 ???? |

Why does the discrepancy of one-buck arise? The conceptual flaw in such a "trick question" appears to be the practical scenario in the bioenergetics field. The overall accounting was erroneously tabulated, owing to the "acceptance of chemiosmosis as the driving principle". Quite simply, accounting is not possible with TMP because it is a "process coincidental variable" of the mOxPhos routine and it is not the ultimate power source of the process outcomes. TMP is an "expression" of the ongoing reaction (energy being spent/produced) and it is not the "used/usable" energy. Therefore, how can we come to terms with all the cynicisms and disparities discussed herein? The only way is to disconnect physiological ATP synthesis from ATP hydrolysis mediated by Complex V. In mitochondrial physiological ATP synthesis, it was estimated that the rate of synthesis versus hydrolysis ratio was 2.4. To explain this "altered equilibrium" (when the equilibrium ratio of hydrolysis:esterification in physiological aqueous environment is 100000:1), it was inferred that the free energy change approaches a zero value when the reaction is done on the surface of Complex V. The simpler explanation is that there must be another principle synthesizing ATP in mitochondria. Since it is established that Complex V is an essential "coupling factor", we should investigate its role for the steady-state, based on the known ATPase role of Complex V, as a source of providing protons to a semi-finished product (say, a negatively charged radical-adduct formed). This is precisely the role that murburn concept based explanation affords to Complex V.

# ITEM 3

Figure C1 shows a brief layout of the supposed ETC scheme staged at the inner mitochondrial membrane. As evident, electrons put into the ETC seem to go in a zig-zag manner, in ones and twos, only to "make water" at Complex IV. Before comprehensively analyzing the feasibility and utility of the prevailing model of ETC, each element of the ETC shall be studied first.

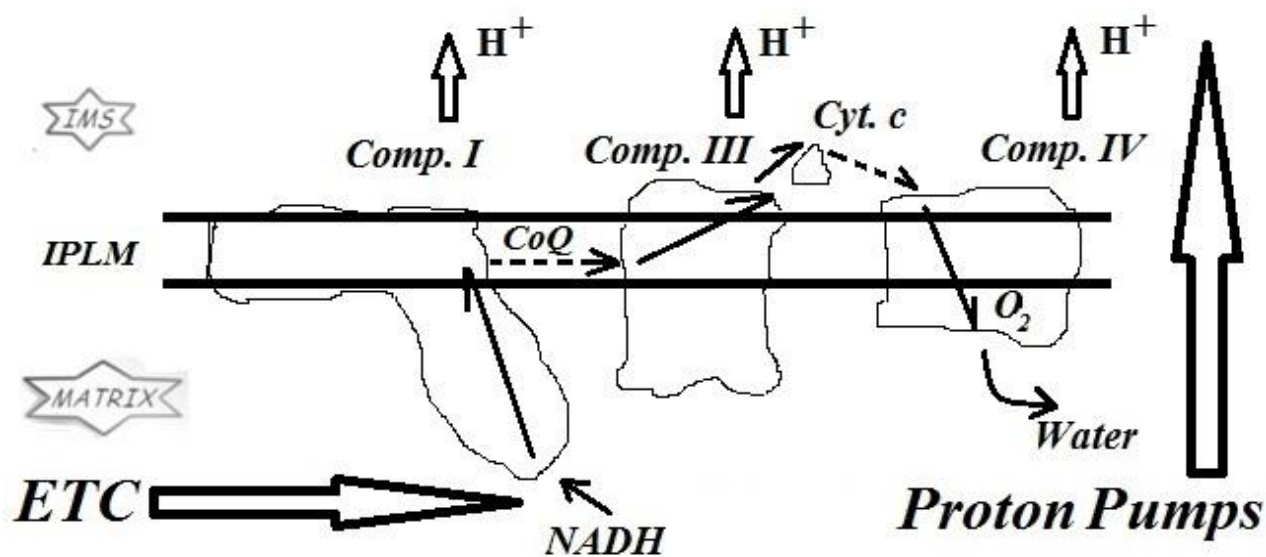

*Figure C1: A simplistic and conservative scheme for the major ETC and proton pumps. (The ET route in respirasome would be quite different, as Complex III would be "perched" within the TM region of Complex I. Also, Complex II is not shown for retaining clarity.) The diagram gives an indication of the various transfers and migrations of one and two electrons (species) involved within and between proteins/complexes. The shapes and sizes of the complexes are crude approximations. The complexes are respectively abbreviated as Comp. IMS & IPLM stand for inter-membrane space and inner phospholipid membrane respectively. While the electron transport is relayed along the inner membrane (along any direction therein), the direction of proton pumps is orthogonal, towards the IMS.*

***Complex I:*** Table C1 (along with a portion of Figure 7 of the main manuscript) depict the overall scheme and analysis of ET (and proton pumping facets) of Complex I.

With a large trans-membrane anchoring portion and a matrix-ward projection, Complex I is a mammoth multi-protein complex with several redox-active centres (Sazanov, 2015).

The redox centers N5 and N6b were found to remain in oxidized state, even under highly optimized "reducing" conditions (Bridges et al., 2012). Table C1 shows that this is because they are flanked by two rather "unfavorable" steps. Further, two Fe-S clusters, N1a and N7 do not lie in the purported "wired" ET route (Sazanov, 2015). Complex I is supposed to serve as a major proton pump, ejecting four protons across the inner membrane, per two electrons received from NADH. Very importantly; the purported ET routes and proton pump loci are highly disconnected. Also, as seen from Table C1, several steps of the proposed route have unfavourable gradients or distances involved.

*Table C1. Analysis of electronic circuitry within Complex I.*

| Step | Donor (mV) | Acceptor / Product (mV) | Distance (Å) | Favorable ΔE? | Favorable distance? | Electron(s) |
|---|---|---|---|---|---|---|
| 1 | NADH (-320) | FMN (-340)$^\$$ | <3? | Probable (high NADH) | Probable | 2e / step (proton required) |
| 2 | FMNH$_2$ (-340)$^\$$ | N3 (-250 to -321); N1a (-233) | 10.9 (7.6); 13.5 (12.3) | Probable; Probable | Probable; Less probable | 1e / step (x 2) |
| 3 | N3 (-250) | N1b (-240 to -420) | 14.2 (11) | Less probable | Less probable | 1e / step (x 2) |
| 4 | N1b (-370) | N4 (-250 to -291) | 13.9 (10.7) | Probable | Less probable | 1e / step (x 2) |
| 5* | N4 (-250) | N5 (-270 to -480); N7 (-314) | 12.2 (8.5); 24.2 (20.5) | Less Probable; Less probable | Probable; Less probable | 1e / step (x 2) |
| 6 | N5 (-430) | N6a (-250 to -325) | 16.9 (14) | Probable | Less probable | 1e / step (x 2) |
| 7* | N6a (-250) | N6b (-188 to -420) | 12.2 (9.4) | Less probable | Probable | 1e / step (x 2) |
| 8 | N6b (-420) | N2 (-50 to -200) | 14.2 (10.5) | Probable | Less probable | 1e / step (x 2) |
| 9 | N2 (-150) | CoQ (-300 to -120) | 11.9 (8.6) | Less probable | Probable | 1e / step (x 2) |

NADH binding was found to approach a value of $10^8$ M$^{-1}$s$^{-1}$ (corresponding to 0.2 μs binding time). [Data for this table were taken from the following publications- ((Bridges et al., 2012; Medvedev et al., 2010; Moser et al., 2006; Sazanov, 2015; Treberg and Brand, 2011; Verkhovskaya et al., 2008)). There seems to be some confusion/mismatch in the nomenclature of the centers (Bridges et al., 2012; Sazanov, 2015). In the manuscript, I have cited N5 and N6b as the "non-oxidized" Fe-S clusters.] *Non-reducible centers; $^\$$Average for 2e process; -389 and -293 respectively for 1$^{st}$ and 2$^{nd}$ e-transfer steps. The values of distance in brackets are the closest edge-to-edge range.

*Complex II:* Quite like Complex I, Complex II has a flavin cofactor and a bevy of Fe-S clusters. In addition, there is also a heme stationed towards the inter-membrane space side, which is not supposed to be a part of the ET route (Sun et al., 2005). The analysis of ET within this complex is given in Table C2. As seen, the transfer step 3 is of low probability. Some researchers opine that this complex has two CoQ binding sites and therefore, this complex may be involved in Q-cycle (Cecchini, 2003). For reasons unknown, this complex is not supposed to pump protons across the inner membrane.

*Table C2: Analysis of electron transfers within Complex II.*

| Step | Donor (mV) | Acceptor / Product (mV) | Distance (Å) | Favorable ΔE? | Favorable distance? | Electron(s) |
|------|------------|-------------------------|--------------|---------------|---------------------|-------------|
| 1 | Succinate (-31) | $FADH_2$ (-79) | 4.6 (2.5) | Probable | Probable | 2e / step |
| 2 | $FADH_2$ (-79) | 2Fe-2S (0) | 16 (11.1) | Probable | Low probability | 1e / step (x 2) |
| 3 | 2Fe-2S (0) | 4Fe-4S (-260) | 12.4 (9.3) | Low probability | Low probability | 1e / step (x 2) |
| 4 | 4Fe-4S (-260) | 3Fe-4S (+60) | 11.9 (9.1) | Probable | Probable | 1e / step (x 2) |
| 5a | 3Fe-4S (+60) | CoQ (+113) | 11 (7.6) | Probable | Probable | 1e / step (x 2) |
| 5b | 3Fe-4S (+60) | Heme b (-185) | 18.5 (11.4) | Low probability | Low probability | 1e / step (x 2) |
| 6 | Heme b (-185) | CoQ (+113) | 6.5 (9.8) | Probable | Probable | 1e / step (x 2) |

The turnover rate of the final step of CoQ reduction was found to be $10^2$ to $10^3$ $s^{-1}$. Data for this table were taken from (Anderson et al., 2005; Sun et al., 2005; Yankovskaya et al., 2003)

*Complex III & Q-cycle:* The cycling of CoQ (or Q-cycle, a junction of circuitry between the electron-equivalents input from NADH via Complex I and from succinate via Complex II) is an integral part of the RCPE proton-centric mechanism. The execution of Q-cycle at Complex III is a highly ingenious mechanism proposed in modern biochemistry (Figure C2). CoQ is the most crucial diffusible component within the inner phospholipid membrane of mitochondria. At Complex III, a Q-cycle requires two fully reduced CoQ, one oxidized CoQ, two protons from the matrix side and two Cyt. *c* from the inter-membrane side (with specifically chartered movements of Fe-S Rieske protein within in a very precisely coordinated time scale). That is, in a single step, three different molecular species are supposed to bind simultaneously to three different sites of Complex III. In the first step of the cycle, two intra-membrane sites Qo (also called P site) and Qi (also called q site) bind to $CoQH_2$ and CoQ respectively and an inter-membrane (Cyt $c_1$) site binds to oxidized Cyt. *c*. This leads to an expulsion of two protons to inter-membrane space and reduction of Cyt. *c* and partial reduction of CoQ at Qi site (leading to the semiquinone). In the second step, a molecule of CoQ and a molecule of oxidized Cyt. *c* bind at Qo and Cyt *c*1 sites respectively (and the Qi site remains occupied with the one-electron reduced, semiquinone form of CoQ). Subsequently, Complex III draws two protons from the matrix (or is assisted by Complex II function in this regard?), and gives the formation/release of reduced $CoQH_2$ and Cyt. *c* at sites Qi and Cyt *c*1 respectively (with concomitant release of CoQ at Qo). Thus, the Q cycle starts with two $CoQH_2$ to generate two separate molecules of reduced Cyt. *c* and regenerates a reduced $CoQH_2$. The scheme, though elegant, is highly fastidious and quite difficult to justify theoretically/practically. Figure 4 of a review by Moser *et al.* shows a clear snapshot of the architecture of redox centres within this complex (Moser et al., 2006). Distances between the redox centres are overwhelmingly large and there is no known logic for ensuring the outcomes within such a fastidious sequence. The "swinging" of the iron-sulfur or Rieske protein and electron/proton "pumping and gating" mechanisms are supposed to achieve the electronic circuitry's "closure" (Crofts et al., 2006). Why or how would the electrons flow from $CoQH_2$ back to CoQ? Rephrased, why should the quinone intermediate(s) give and take electrons in the same cycle, at the same Complex? How is the molecular intelligence of such a "deterministic" scheme effected? Such a scheme would not have any thermodynamic drive and these events would be low on probability. If this fastidious Q-cycle fails, the whole ETC circuitry breaks down. If the recently determined respirasome structure (Wu et al., 2016) has real physiological significance (as it must!), the CoQ cycle becomes a redundant facet, thereby seriously undermining the hitherto perceived wiring concept of ETC. Quite curiously, the bulbous matrix-ward extension of this complex does not possess any functional relevance, as per the RCPE scheme. Complex III was analyzed with the ET flow starting with the step

CoQH$_2$-FeS for both loops because the CoQH$_2$ to Cyt $b_L$ distance was larger than 12 Angstroms (edge to edge) and there is little reason for an ET split at the first/binding step. [Source for redox potentials within Complex III: Anthony Crofts webpage at UIUC- http://www.life.illinois.edu/crofts/bc-complex_site/ & (Iwata et al., 1998; Zhang et al., 1998).] The distances between the respective hemes and FeS centres in the dimers are from 21 to 63 Angstroms apart and therefore, they are not considered relevant for intermolecular ET phenomena. The data analysis is stemmed on the belief system that ubiquinone (reduced and oxidized) interact at two locale on the enzyme- the reduced species interacts via the Fe-S protein on the inter-membrane side and the oxidized species interacts via the Cyt $b_H$ on the matrix side. The FeS Rieske protein is deemed a bifurcating point, with options to give the electron to Cyt $c_1$ in the inter-membrane space or Cyt $b_L$ towards the matrix side of the inner membrane. The last step electron transfer timescale (to Cyt $c$) is in the range of $10^0$ to $10^2$ μs (the fastest step of the overall process) and each of the other steps may incur a time window of $10^2$ to $10^3$ μs.]

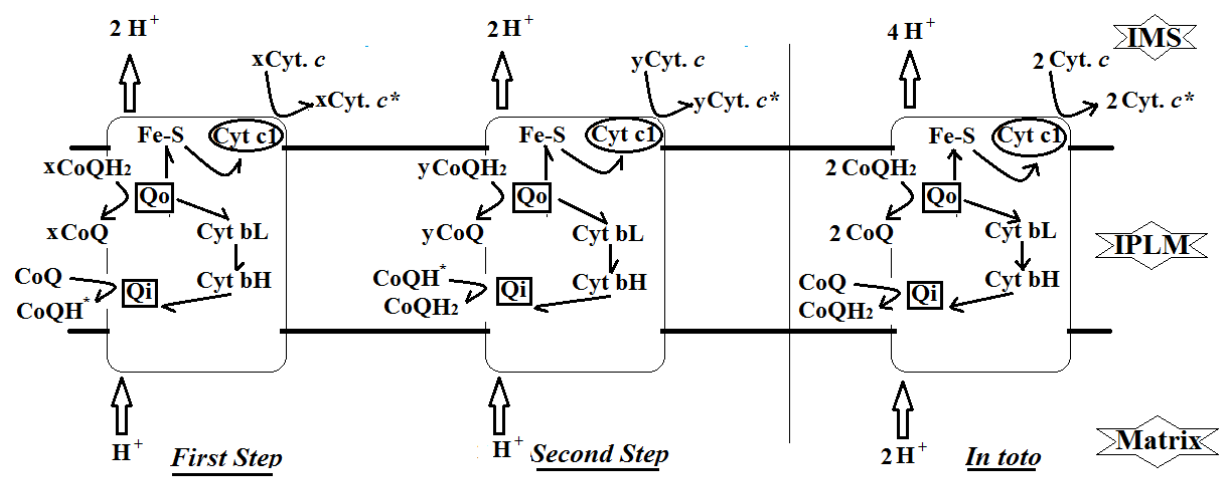

*Figure C2: The various steps and circuitry of Q cycle transpiring at Complex III. Detailed description is provided within the text above. The two CoQ binding sites are marked in square boxes and the Cyt. c binding site is marked by ellipse. One of the Fe-S clusters is not depicted for the sake of clarity. The x and y prefixes for CoQ and Cyt. c connote two different molecules (not numbers).*

***Complex IV:*** The last pit-stop of the ETC is the Cu + pentacoordinated heme containing Complex IV. The rate of anaerobic intra-molecular electron transfer between the two hemes ranged from 0.02 to 7 per second, which is several orders lower than the overall experimental electron transfer rates (Orii, 1988). This is the instance of a simple one-electron transfer step within the last leg of ETC, well within the "theoretically favorable" criteria (permissible distance and redox potential gradient). Therefore, ET based on the "outer sphere model" does not explain electron transfers between redox centres even "within" this pivotal protein. It must be noted that the overall ETC becomes defunct if the oxygen's bound presence within Complex IV is not a highly fecund and tight binding process. Therefore, here we need to contend with the supposition that molecular oxygen binds to the heme-Cu centre of Complex IV and stays committed to it for a protracted time frame, to sequentially receive four protons and four electrons, before dissociating as two molecules of water. It remains a conundrum why the intermediary species should not dissociate, leading to DROS formation. It is well-known that binding of a diatomic gaseous molecule like oxygen to a metal centre is a temporal "on-off" (binding-detachment) process dictated by thermodynamic equilibrium. Usually, at low ligand concentrations, the amount of time the ligand is "off" is greater than the amount of time it is "on". Therefore, it is inconceivable that a molecular tethering mechanism could evolve or be applicable for all of the different species (the combinations of diatomic oxygen ± one to four electrons ± one to four protons) that could potentially be formed on that reaction centre. Even more interesting is to note the fact that while the $K_d$ for oxygen was found to range between the values of 0.3 to 1 mM, the $K_M$ value was several orders lower, at ~1 μM. This is when theory dictates that $K_M \geq K_d$. [as $K_M = K_d + (k_2/k_1)$.] Classical Michaelis-Menten binding-based kinetics treatments cannot explain this outcome. Further, with a $K_d$ value of 0.1 to 1 mM for Complex IV-$O_2$ interaction, the $k_{off}$ would range from $10^4$ to $10^6$ per second (assuming a diffusion limited binding value, ranging from $10^8$ to $10^9$ $M^{-1}s^{-1}$). It is very difficult to envisage that such an inefficient binding process would ever give an overall oxygen reduction rate of ~$10^3$ per second (which happens to be the physiological rate of oxygen reduction). It is also difficult to envisage how four protons are relayed into the phospholipid membrane (where the reaction centre is located). [For Complex IV, the source for redox potentials was – (Moser et al., 2006) and hyperphysics webpage- http://hyperphysics.phy-astr.gsu.edu/hbase/Chemical/redoxp.html]

*Analysis of the overall kinetics of ETC:* The thermodynamic drive for electron flow/transfers in mitochondrial ETC scheme cannot be a push of electron(s) from lower to higher redox potential proteins because the arrangements of several redox centres within the ETC flout the requisites (of a gradual increment of the participating elements' redox potentials). Further, as cited earlier, our group's findings with heme-flavin proteins and small molecules have challenged such notions. The thermodynamic drive cannot be a pull either because oxygen bound at Complex IV cannot exert such an effect via the diffusible species of Cyt. *c*, through the membrane-bound Complex III, and through the membrane-soluble CoQ, all the way until Complexes I or II. Therefore, without a significant push or pull operating, the electron transfer rate equation [log ($k$) = 12.8 – 0.6 d] applies; where *k* is the rate (per second) and d is the distance in Å (Moser et al., 2006). Since the overall water formation rate has been experimentally noted to be >$10^3$ per second (Orii, 1988), the sequential ETC scheme solicits that most transfers must be done at least at frequencies of $10^5$ to $10^6$ $s^{-1}$. For the same, the maximal permitted distance corresponds to ~12 Å. This value approximates the distance of ~15 Å, the maximum limit perceived for efficient outer-sphere electron transfers in most proteins (Canters and Vijgenboom, 2012; Da Silva and Williams, 2001). Since the physiological proton concentration is ~$10^{-7}$ M, the reorganization time for most proteins must fall in the range of $10^{-3}$ to $10^{-2}$ seconds (depending on the constraints protons experience to access the pertinent moiety on the protein). Now, these realities (as summated above) pose insurmountable constraints on the sequential mode of a bevy of electron transfers supposed to occur in the ETC.

It is known and generally agreed that transfer of a single electron between two redox centers (of the various combinations like- flavin→FeS, FeS↔FeS, FeS↔CoQ, FeS→heme, CoQ↔heme, heme↔heme, heme↔Cu, etc.) across favorable potentials and permissible distances would be micro- to milli- second scales phenomena. For example, the electron transfer within Complex I (from the FMN, all the way to the last Fe-S center, N2) was experimentally found to occur in ~$10^2$ μs range (Verkhovskaya et al., 2008). It can be safely inferred that an elaborate and sequential ETC circuitry with individual components such as the above, and the example of heme-heme transfer within Complex IV already cited, cannot afford the overall experimental ET rate observed in functional systems.

*Conceptual notes:* Without going into the intricacies of Marcus' theory on ET, a simple understanding of chemical principles vouches that one-electron donating and accepting ability within the mitochondrial regime would depend primarily upon the key chemico-physical attributes of - species' redox potentials, concentrations, mobility/distances, partitioning and stability of the participating/resulting entities, etc. Though electrons do get transferred against redox potential gradients in solution chemistry, it is only viable when the concentrations permit them to. If the entity with a lower redox potential should receive an electron from a higher potential species, the concentration of the latter should be several folds or orders higher than the former. The only way this can be obviated is when the donor-acceptor pairs are immobilized and we apply an external potential to drive the reverse-gradient transfer. In the purported mitochondrial ETC, the counter-gradient transfer is not possible within a protein because each redox couplet (donor-acceptor pair in the "wired" sequence) is at 1:1 ratio. Further, several researchers perceive the redox centers' connectivity within the proteins as a "wire" (as exemplified by the linear connectivity drawn within matrix-ward projections of Complexes I & II). This might be a misplaced perception because a wire "conducts electrons" when/where there is a potential difference. If there is no applied field, no electrons would flow through a metal wire. Potentials are created or exist, when electrons or negative/positive charges accumulate at a given point. Clearly, the wire-analogy is inapplicable here.

Furthermore, in biological systems, there cannot be "freely-transferable high or low energy electrons or protons". An electron in a particular orbital of an atom/molecule may have higher/equal energy term when compared to/with another electron in another orbital within the same/similar atom/molecule. But, if the electrons are mobile, they should be seen as "similar". In other words, when an electron (or proton) changes its "address", it retains little information regarding its previous "residence". The events transpire under normal physiological conditions and there is no external heat (to enhance its thermal energy) or electric field input (to accelerate it). This argument is very important because in several cases, an electron at certain junctions of the "wired ETC" can move further only by going against the "resting potential" and they seem to "somehow upgrade their stature" at their new locus. Therefore, the diagrams that show ATP synthesis occurs at "energy differentials" (when electrons are transferred through two definite carriers within such a "roller-coaster" of an ETC) are theoretically incorrect.

The above statements need to be rearticulated again. How could an electron move from a lower to an upper rung of potential via a junction, suddenly gain energy at the higher energy locus and yet remain directionally mobile? If it is freely mobile, energy profile cannot be deemed important. If energy profile is important, it cannot move freely against gradient in a connected system. Comparing electron flow with water flow is erroneous in the context because water can log at a point (because of a dip in the contour) and the water column at a given locus may also gain height owing to capillarity. In the proteins, there is no way for several electrons to reside at a junction (similar to water-clogging) and thereby raise the electron's potential at that locus to a higher status. The junctions in context are one-electron donors/receivers and the next stop in the relay is usually too far for an electron to "tunnel" through. Also, it is a very constrained aesthetic and deterministic notion that electrons would be transferred in pairs or only across narrow potential differences, as the prevailing ETC concept seeks. With these fundamentals in place, the mitochondrial ETC is further analyzed at a comprehensive scale.

***Overall connectivity issues:*** Modularity, ordered arrangements or synchronization mechanisms are not evident in the mitochondria. Further, since the membrane is a mosaic fluid with a "delocalized relative positioning" of all participating components, it is very difficult to imagine a "high-fidelity circuitry" that the RCPE paradigm solicits. While NADH and succinate could transiently deliver two electrons to the Complexes I and II systems via the bound flavin cofactors, it remains an enigma as to how CoQ receives and gives two one-electron equivalents in its interactions with FeS Complexes (in Complexes I/II and Complex III respectively). For justifying the intricate ETC within Complex III, the incoming electron pair from $CoQH_2$ must be spontaneously split and parted to an Fe-S cluster and a heme-center. This is when the purported CoQ binding site (Qo) is ~6.5 and ~12.5 Angstroms (edge-edge) away from the nearest/respective Fe-S and heme-centers (Moser et al., 2006), and the heme-center does not pose a favourable potential gradient for electron transfer. Such requisites are highly fastidious and practically improbable to achieve. Anytime this switch does not work, the overall sequential ETC would break down. Why should Complex III route electrons through the one-electron agent of Cyt. *c*, and that too through another aqueous phase? Surely, this would impede the electron transfer efficiency to Complex IV. How can so many redox active molecules and centers stay dedicated to their designated roles, without the circuitry getting shunted and without liberating DROS? It would have served well if the protein had only one or two redox centers. Why should there be so many redox centers within the protein complexes and why are multiple one-electron transfers seen within a protein complex? These facets would surely increase the probability of DROS formation in the system. How could the ETC ever function in the presence of oxygen and DROS? Why electron transfer routes are so disconnected from the purported trans-membrane proton pumps motifs and how do these get to work in tandem? How can the passage of one-electron through a path within the circuit pump two protons out? How can the electron transfers that occur in microsecond timescales be coupled to trans-membrane proton pumps that occur in millisecond timescales?

***Summating perspectives on proton-centric ETC*:** The proton-centric explanations do not offer any convincing rationale for the criss-cross movement of single or pairs of electrons (Figure C1). Also, if we indulge the premise that the proton-pump supposition is invalid (Manoj, 2017), why should electrons go through a circuitous (deterministic) route for merely reducing $O_2$ bound at Complex IV? Fundamentals of reaction chemistry dictate that super-coordinated multi-molecular sequential reactions would have low probabilities to occur spontaneously and repeatedly. As discussed, such reactions cannot add up to overall conductions within micro/milli- second time frames, particularly within the physiological concentrations of reactants, limited mobility and spatial constraints posed by the phospholipid environment. The current ETC concept does not have any explanation for the three "non-functional" redox centers seen in Complexes I/II and cannot explain the non-reducibility of redox centers in Complex I. The RCPE concept does not explain the fundamental observation of electron transfer in metabolic state 2 (without ADP+Pi) and the elevation of rates in metabolic state 3 (with the addition of ADP+Pi), as seen in experimental mitochondrial research practice. The ETC was supposed to prevent the formation of ROS, but it is seen in plenty within mitochondria. Therefore, the prevailing ETC scheme does not serve mechanics, aesthetics and kinetics. On one end, the RCPE hypothesis seeks the macroscopic perspective of physical separation of protons and electrons in space and time. At the other end, it requires the very protons and electrons to be drawn together with great abundance and absolute accuracy/precision at certain defined loci/times alone. By any thought, for such an ETC to be operational in the physiological time frames, significant energy must be expended.

# ITEM 4

The RCPE explanation for mOxPhos hinges on the spatial and temporal separation (over relatively large distance and time frames, across distinct macroscopic phases) of protons and electrons for forming the electro-chemical trans-membrane gradient (chemiosmosis principle). Further, the membrane-based electronic circuitry and proton-motive motor are supposed to work as highly synchronized distinct modules, in conjunction with the Krebs' cycle and some other bioenergetic pathways. As shown in Figure D1, the prevailing RCPE paradigm could be compared with an automobile's functional elements for energetics or a hydroelecic power plant (dam) for its structural elements (minus the "ETC + proton pumps" components). When comparing the chemiosmotic model of mitochondria with working man-made machine like automobile, the presence of following functional modular entities are solicited-

(i) bio-engine (ETC: burn the fuel NADH to generate the energy to do the work, retain difference of electron accepting-donating redox couples across a complex, contiguous and compartmentalized "organic" circuitry),

(ii) bio-dynamo (Proton pumps: generate a proton-electro-chemical gradient across the inner membrane),

(iii) bio-battery (ATPsynthase: serve as the mechano-energetic coupler to cyclically synthesize the energy currency, which possesses the ability to do chemical work), &

(iv) bio-sensor & bio-regulator/pacemaker (Chemiosmosis logic: self-analyze and self-regulate proton concentration, movements of electrons, tap into electrical field, etc. and thereafter, govern the movement/synchronization of molecular motor).

*Figure D1: Pictorial renditions of working analogies of RCPE hypothesis for mOxPhos. IMS and IPLM stand for inter-membrane space and inner phospholipid membrane respectively. In the first analogy on the left, protons can be analogous to the role played by water in a hydroelectric power plant (dam). In the analogy on the right, the ability to*

*utilize/capture electro-chemical energy and use it for chemico-mechanical work is shown. (In both images, factual components are shown in normal fonts whereas the analogy is italicized.)*

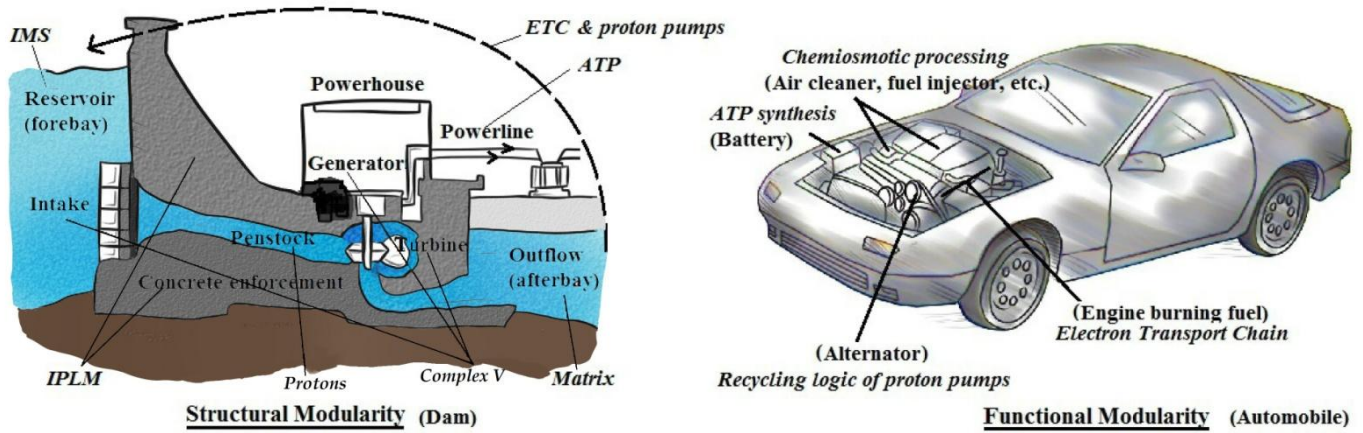

As evident from the analogy, it is highly unlikely that such a working logic could evolve from a minimal set of components. Quite simply, such a working logic can only be assembled or created.

In stark contrast, the newly proposed murburn explanation for mOxPhos is a decentralized ATP synthesis route by a one-electron scheme. In most constitutive metabolic reaction schemes/cycles (of the type A→B→C→D…→X), the major drive for the formation of X is the accumulation of A, B, C, etc. within the reaction milieu/pool. (This is a tentative scenario assuming the lack of feedback inhibitions and allosteric regulations.) Therefore, the protagonists formed in the initial part exerting a forward push forms the major thermodynamic drive for the overall reaction. Quite differently, in the mXM system, it was shown that the two-electron reacted products formed at the last reaction stage (formation of X) exerts a thermodynamic pull on the overall one-electron equilibriums (in the initial part of the reaction cycle) to shift towards right. The reaction sequence does not seek compartmentalization or ordered/ sequential events. It was already demonstrated that such a one-electron process could be mild, reproducible, specific/selective and inherently constitutive in nature. The overall scheme affords a viable redox chemical logic in the physiological regime, one which does not seek affinity-based ter-/multi-molecular interactions. In stark contrast to the analogy with an automobile/hydroelectric project per the erstwhile RCPE hypothesis, the newly proposed system is analogous to a simple hearth with respect to structural modularity or nuclear fission reactor in terms of functional logic (Manoj, 2017) as shown in Figure D2. The latter system can form spontaneously and does not need intelligent controls from the outside, but is capable of being regulated from within.

*Figure D2. A comparison of murburn concept of mOxPhos with some simple setups*

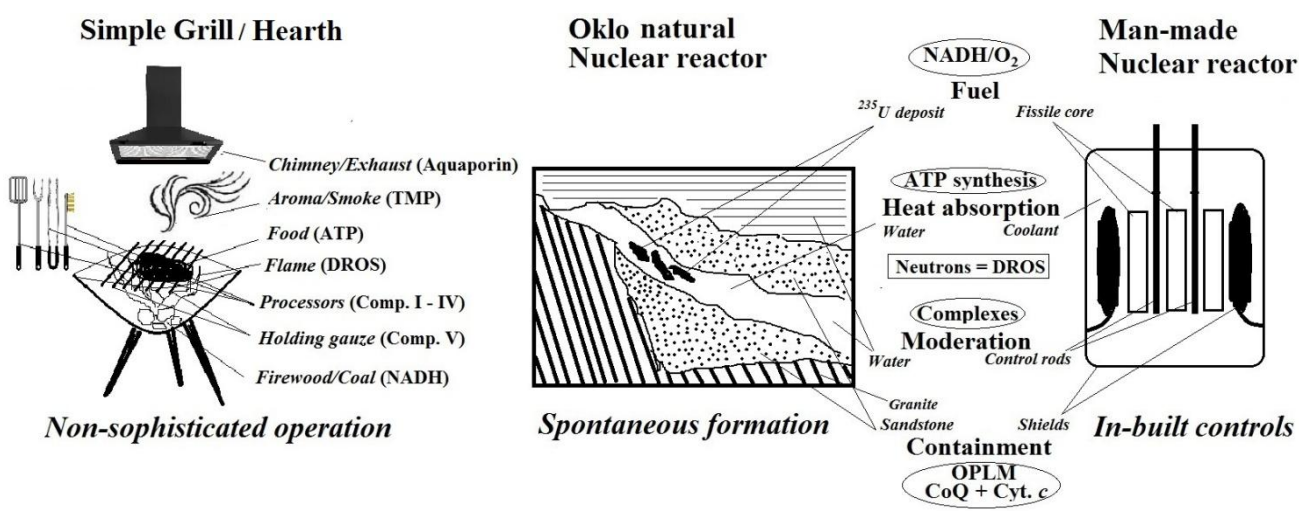